\documentclass[twocolumn,english]{llncs}
\usepackage[T1]{fontenc}
\usepackage{babel}
\usepackage{amstext}
\usepackage{graphicx}
\usepackage[unicode=true,
 bookmarks=true,bookmarksnumbered=true,bookmarksopen=true,bookmarksopenlevel=1,
 breaklinks=false,pdfborder={0 0 0},backref=false,colorlinks=false]
 {hyperref}
\hypersetup{pdftitle={Your Title},
 pdfauthor={Your Name},
 pdfpagelayout=OneColumn, pdfnewwindow=true, pdfstartview=XYZ, plainpages=false}

\makeatletter

\usepackage{refcount}
\usepackage[noresetcount]{algorithm2e} 

\IncMargin{1em}
\LinesNumbered
\SetAlCapSkip{0.5em}
\usepackage{todonotes}

\usepackage[margin=1in]{geometry}

\makeatother

\begin{document}

\title{Dynamic concurrent \\
van Emde Boas array}

\author{Konrad Ku\l akowski}

\institute{AGH University of Science and Technology, \\
al. Mickiewicza 30, Kraków, Poland, \\
\href{mailto:konrad.kulakowski@agh.edu.pl}{konrad.kulakowski@agh.edu.pl}}
\maketitle
\begin{abstract}
The growing popularity of shared-memory multiprocessor machines has
caused significant changes in the design of concurrent software. In
this approach, the concurrently running threads communicate and synchronize
with each other through data structures in shared memory. Hence, the
efficiency of these structures is essential for the performance of
concurrent applications. The need to find new concurrent data structures
prompted the author some time ago to propose the cvEB array modeled
on the van Emde Boas Tree structure as a dynamic set alternative. 

This paper describes an improved version of that structure - the dcvEB
array (Dynamic Concurrent van Emde Boas Array). One of the improvements
involves memory usage optimization. This enhancement required the
design of a tree which grows and shrinks at both: the top (root) and
the bottom (leaves) level. Another enhancement concerns the successor
(and predecessor) search strategy. The tests performed seem to confirm
the high performance of the dcvEB array. They are especially visible
when the range of keys is significantly larger than the number of
elements in the collection.
\end{abstract}

\section{Introduction\label{sec:Introduction}}

The rapid rise in the popularity of multi-core shared-memory processor
systems makes concurrent programs increasingly common and desirable.
Following the growing market for concurrent software, an increasing
demand for the use of concurrent data structures can be observed.
Such a situation makes the search for new concurrent data structures
particularly important. One of the attempts to find such a structure
is the work \cite{Kulakowski2014acve} in which the author proposed
the early version of the \emph{concurrent van Emde Boas} (\emph{cvEB})
array. The structure presented here is an example of concurrent dynamic
set implementation providing, in addition to the standard methods
\emph{insert()}, \emph{remove()} and \emph{find()}, also the method
\emph{successor()}, which allows users to determine the first greater
element from the specified one. It has very good theoretical and practical
properties, as confirmed by tests and analyses carried out. Unfortunately,
one of the shortcomings of that solution is the need to allocate all
the required memory at the very beginning, as in the case of a regular
array. Another limitation of that structure is the implementation
of \emph{successor()}, which in the case of massive interference with
\emph{remove()} operating in different threads might be delayed or
failed due to the search repetition. These deficiencies led the author
to propose a new \emph{dynamic concurrent van Emde Boas} (\emph{dcvEB})
array, which, on the one hand, retains the good properties of its
antecedent, and on the other hand is deprived of its shortcomings.
Hence, the new structure presented in this article allocates and deallocates
memory dynamically, depending on the amount of data stored in it.
In addition, a new strategy for the \emph{successor()} and \emph{predecessor()}
methods has been adopted. The new structure, rather than repeating
the successor or predecessor search, continues searching until an
appropriate element is found or the absence of such an element is
decided. The assumed strategy is more robust and less susceptible
to interference. It also seems to be more intuitive and justifiable
in the context of user expectation. 

The article consists of several sections, where, except for introductory
ones (Sec. \ref{sec:Introduction} - \ref{sec:van-Emde-Boas}), the
\emph{dcvEB array} (Subsection \ref{sub:Structure-organization})
and its implementation (Subsection \ref{sub:dcvEB-array-methods})
are discussed. Next, the mechanisms of concurrent expanding and shrinking
are explained (Subsection \ref{sub:Expanding-and-trimming}). Other
enhancements, such as dynamic memory allocation and the new search
strategy, are explained (Subsections: \ref{sub:Dynamic-adding-and}
and \ref{sub:Successor-search-strategy}). Then, the successor search
running time and the structure correctness are discussed (Section
\ref{sec:Concurrency-and-dynamism}). The experimental results are
examined in Section \ref{sec:Experimental-results}. The comments
and discussion (Section \ref{sec:Comments-and-discussion}) and a
brief summary (Section \ref{sec:Summary}) close the article.

\section{Background}

A dynamic set is one of the basic data structures in computer science.
Usually, it is assumed that a dynamic set supports the following operations:
\emph{insert()}, \emph{delete()},\emph{ search()}, \emph{minimum()},
\emph{maximum()}, \emph{successor()} and \emph{predecessor()} \cite[p. 230]{Cormen2009ita}.
The first two of them are included in the category of \emph{modifying
operations}, while others are \emph{queries}, which do not modify
the structure.  Due to increasing demands for data format, different
structures support dynamic set operations to varying degrees. In particular,
good dynamic set operation performance is provided by\emph{ balanced
search trees}. For instance, all the dynamic set operations can be
handled by \emph{RB-Trees} \cite{Bayer1972sbbtSIMPL} in a sequential
running time $O(\lg\alpha)$, whilst \emph{van Emde Boas} trees \cite{Boas1977poia}
need barely $O(\lg\lg\alpha)$ time to complete any of the mentioned
operations \cite{Cormen2009ita}. Unfortunately, transition from the
sequential to the concurrent objects is not easy \cite{Shavit2011dsit}.
Hence, many concurrent dynamic set implementations (e.g. \cite{Ellen2010nbbsSIMPL,Howley2012anbiSIMPL})
do not support all the dynamic set operations and instead focus on
dictionary operations. 

The early works on the concurrent \emph{balanced search trees} with
\emph{dictionary} operations began to emerge in the 70s \cite{Samadi1976btia,Bayer1977coooSIMPL}.
In the subsequent years, the topic was studied in \cite{ELLIS1980csai,Kung1980cmobSIMPL,Lehman1981hcSIMPL}.
The studies, initially focusing on lock strategy \cite{Bayer1977coooSIMPL}
and lock coupling \cite{Lehman1981hcSIMPL,Lanin1986ascbSIMPL}, began
to deal with the relaxed (delayed) re-balancing \cite{Nurmi1996ratmSIMPL,Hanke1997rbrb}
and the non-blocking synchronization schemes \cite{Bronson2010bsSIMPL,Ellen2010nbbsSIMPL,Brown2011nbksSIMPL,Howley2012anbiSIMPL}. 

\emph{Skip List}, proposed by \emph{Pugh} \cite{Pugh1990cmos}, is
an alternative to \emph{balanced search trees}. It provides several
linked lists arranged in a hierarchy, so that the single list corresponds
to the set of nodes at the same depth in a \emph{search tree}. The
structure avoids additional re-balancing due to the randomized fashion
of the insertion algorithm. \emph{SkipList }is suitable for both the
sequential and concurrent applications. Very efficient \emph{SkipList}
implementation \cite{Herlihy2007asos}, based on \emph{Fraser} \cite{Fraser2004plf},
is part of a standard \emph{Java API} \emph{5}. The \emph{Java} \emph{SkipList}
implementation as one of the few (the second is a \emph{SnapTree}
\emph{Map} by \emph{Bronson} \cite{Bronson2010bsSIMPL}) supports
all the dynamic set operations including \emph{successor() }and \emph{predecessor().}

\section{van Emde Boas tree \label{sec:van-Emde-Boas}}

The tree structure proposed by \emph{van Emde Boas} \cite{Boas1977poia}
is not a typical search tree. It supports all the dynamic set operations,
such as \emph{insert()}, \emph{delete()},\emph{ search()}, \emph{minimum()},
\emph{maximum()}, \emph{successor()} and \emph{predecessor()} \cite[p. 230]{Cormen2009ita}
in $O(\ln\ln\alpha)$. This tremendous speed involves the requirement
that the keys must be unique integers in the range 0 to $\alpha-1$.
Thus, from a practical point of view, the \emph{van Emde Boas} \emph{(vEB)
tree} is something between an array and a search tree. Assuming that
the number of stored elements is essentially smaller than $\alpha$,
the \emph{vEB }tree is better than the array as regards the speed
of \emph{successor(),} \emph{predecessor()}, \emph{minimum()} and
\emph{maximum()}. 

Of course, the efficiency of the array operations \emph{insert()},
\emph{delete()} and \emph{search()} remains unchallenged regardless
of the stored data size. In general, the \emph{vEB tree} operates
faster than the other search trees. However, the strong constraint
on the key values makes it unusable if the stored objects cannot be
represented as unique integers. 

The key to the efficiency of the \emph{vEB tree} operations is the
uneven number of subtrees on different levels of the \emph{vEB tree}.
Thus, the root node has $\alpha^{1/2}$ of subtrees, whereas each
next level of \emph{the vEB tree} shrinks the number of children in
the nodes by the square root. Assuming that an operation over the
\emph{vEB tree} performs $O(1)$ work at each level of the hierarchy,
the running time of a method is $O(h)$, where $h$ is the height
of the \emph{vEB tree}. Reducing the number of subtrees can not be
carried out indefinitely. Thus, at the last but one level of the tree,
the nodes have at most two single-element subtrees, i.e. $\alpha^{1/2^{h-1}}=2$.
Hence, we obtain $\ln\alpha=2^{h-1}$, and finally $h=\ln\ln\alpha+1$.
Thus, the asymptotic running time of an exemplary operation is $O(\ln\ln\alpha+1)=O(\ln\ln\alpha)$. 

To be able to traverse each level of a tree in $O(1)$ the \emph{vEB
tree} methods use the arrays of references to the subtrees. For this
reason the root node $T_{\textit{root}}$ needs to store $T_{\textit{root}}.\textit{arr}$
- $\alpha^{1/2}$-element array of subtrees, their children $T$,
$T.\textit{arr}$ - $\alpha^{1/4}$-element arrays of their subtrees,
and so on. With this construction, every method can calculate in which
subtree the given key can be found. For example, in the case of the
root node, the key $x$ is expected to be in $\left\lfloor x/\alpha^{1/2}\right\rfloor $
subtree etc. 

To achieve $O(1)$ level traversing time, the more complex methods
like \emph{successor()} and \emph{predecessor() }need further information
about the subtrees. Thus, with every node $T$ the next three variables
are assigned: $T.\textit{max}$, $T.\textit{min}$ and $T.\textit{summary}$,
where $T.\textit{max}$, $T.\textit{min}$ denote correspondingly
the maximal and the minimal value of a key in the subtree rooted in
$T$. The \emph{summary} is an auxiliary search structure. Intuitively
speaking, the \emph{search()} method traverses down the \emph{vEB}
tree along a well-defined path from the root to the given key. The
\emph{successor()} must deviate from this path to the right (\emph{predecessor()}
to the left). The decision whether to go down into the subtree according
to the predetermined path or go to the right at the same level is
taken on the basis of the value $T.\textit{max}$. Thus, if $T$ is
a subtree in which, according to the path calculation, the key $x$
should be stored, then the \emph{successor()} goes down into $T$
only if $x<T.\textit{max}$, i.e. when the maximal key in $T$ is
greater than $x$. If $x\geq T.\textit{max}$ the \emph{successor()
}method needs to move horizontally to the right in search of the first
non-empty subtree. Of course, such a horizontal search might be time
consuming. For instance, the linear browsing $T_{\textit{root}}.\textit{arr}$
may take up to $O(\alpha^{1/2})$. In order to shorten the horizontal
search, the same mechanism as in the case of the whole structure is
used. $T.\textit{summary}$ is an auxiliary tree that holds information
about the occupancy of the array $T.\textit{arr}$ in the same manner
as the main tree holds the keys. Thus, traversing $T.\textit{summary}$
takes at most $O(\ln\ln\left|T.\textit{arr}\right|)$. In the results,
the overall asymptotic running time of \emph{successor()} and \emph{predecessor()}
is $O(\ln\ln\alpha)$. 

A good and systematic introduction into the \emph{vEB trees} theory
can be found in \cite{Cormen2009ita}.

\section{Construction of the dcvEB array}

\subsection{From the vEB Tree to the dcvEB array\label{sub:From-vEB-Tree}}

One of the reasons why \emph{vEB trees} are not so popular in practice
are space requirements \cite{Cormen2009ita}. The need to allocate
one continuous block of memory in the root of a structure capable
of holding $\alpha^{1/2}$- element array might be inconvenient. The
problem can be addressed in different ways \cite{Mehlhorn1990bodi,Cormen2009ita}.
One of them implemented in the \emph{dcvEB array }proposes the use
of a fixed number of subtrees per node. It results in a worse theoretical
time complexity, however, in many practical applications the achieved
speed appears to be quite sufficient. For the same reason, the \emph{summary}
structure is simplified to a bit-vector aligned to the length of a
machine-word. The use of high-speed non-blocking bitwise operations
on the \emph{summary} vector allows users to avoid the use of $T.\textit{min}$
and $T.max$. The logic behind some methods of the \emph{dcvEB array}
is also changed. For example, in the \emph{vEB tree,} the \emph{delete()}
method performs one single pass from the top to bottom. Due to synchronization
issues in the \emph{dcvEB array} \cite{Kulakowski2014acve} the \emph{delete()}
method proceeds bottom-up. Similarly, \emph{successor()} and \emph{predecessor()}
first reach the bottom of the tree, then start to traverse the tree
moving up and down in search of the appropriate element. The \emph{dcvEB
array} tries to use the non-blocking synchronization mechanisms as
often as possible. For example, the \emph{get()} method uses only
the \emph{lock-free} synchronization mechanisms, which results in
its very good performance in the tests (Sec. \ref{sec:Experimental-results}).
The only exception is the mutual synchronization of \emph{insert()}
and \emph{delete(). }In this case, in order to ensure data consistency
\cite[p. 373]{Kulakowski2014acve} the \emph{readers-writer lock}
\cite{Herlihy2008taom} is used. 

Despite the fact that the creation of the \emph{dcvEB array} was inspired
by the \emph{vEB tree,} the differences between these two structures
seem to be fundamental. Therefore the \emph{dcvEB array} should be
treated, not as a concurrent extension of the sequential \emph{vEB
tree} but, as the new and original data structure.

\subsection{Structure organization\label{sub:Structure-organization}}

The \emph{dcvEB} and cvEB arrays can be seen as a tree of arrays \cite{Kulakowski2014acve}.
Each array's cell holds the reference to \emph{ArrayHolder} \emph{(AH)}
- a tree node structure, which wraps the lower-level array or stores
a specific value if \emph{AH} is a leaf. The leaves are kept at the
lower level of the tree. Each array corresponds to an associated \emph{summary}
- a bit vector, in which the $i-th$ bit is enabled only if the appropriate
array's cell holds the lower level \emph{AH}.  Besides an array and
the associated bit vector, every \emph{AH} also contains the \emph{readers-writers}
lock object \cite{Herlihy2008taom}. The leaf $AH$ instead of an
array reference holds an element and an integer index value as its
key. The value of the key determines the path from the root to the
leaf understood as a sequence of positions on the various levels of
the tree. The path positions are calculated according to the following
recurrent formula: $i_{k}=i_{k-1}-lp_{k-1}*n^{h-k-1},\,lp{}_{k}=\left\lfloor i_{k}/n^{h-k-1}\right\rfloor $
where $n$ is the length of a bit vector, $h$ - is the height of
the tree, $lp_{k}$ is the path position on the $k-th$ level, $i_{0}$
is a key of the element. A position at the root level $lp_{0}$ is
defined as $lp_{0}=\left\lfloor i_{0}/n^{h-1}\right\rfloor $. The
\emph{dcvEB array} of the height $h$ can hold elements within the
range $[0,\ldots,n^{h}-1]$. If there is a need to store an element
with a key greater than $n^{h}-1$ or by removing the item there are
no elements with the keys within the range $[n^{h-1},\ldots,n^{h}-1]$
the tree has to be vertically resized. The concurrent tree resizing
algorithms as integral parts of the \emph{insert()} and \emph{remove()}
procedures are discussed later. If the \emph{dcvEB array} does not
contain a particular element, and its key fits the current key range,
the insert procedure recreates the missing $AH$ along the path from
the \emph{root} to the leaf. Similarly, the remove procedure deletes
$AH$s if the appropriated \emph{summaries} are $0$.

\begin{figure}[h]
\begin{centering}
\includegraphics[scale=0.62]{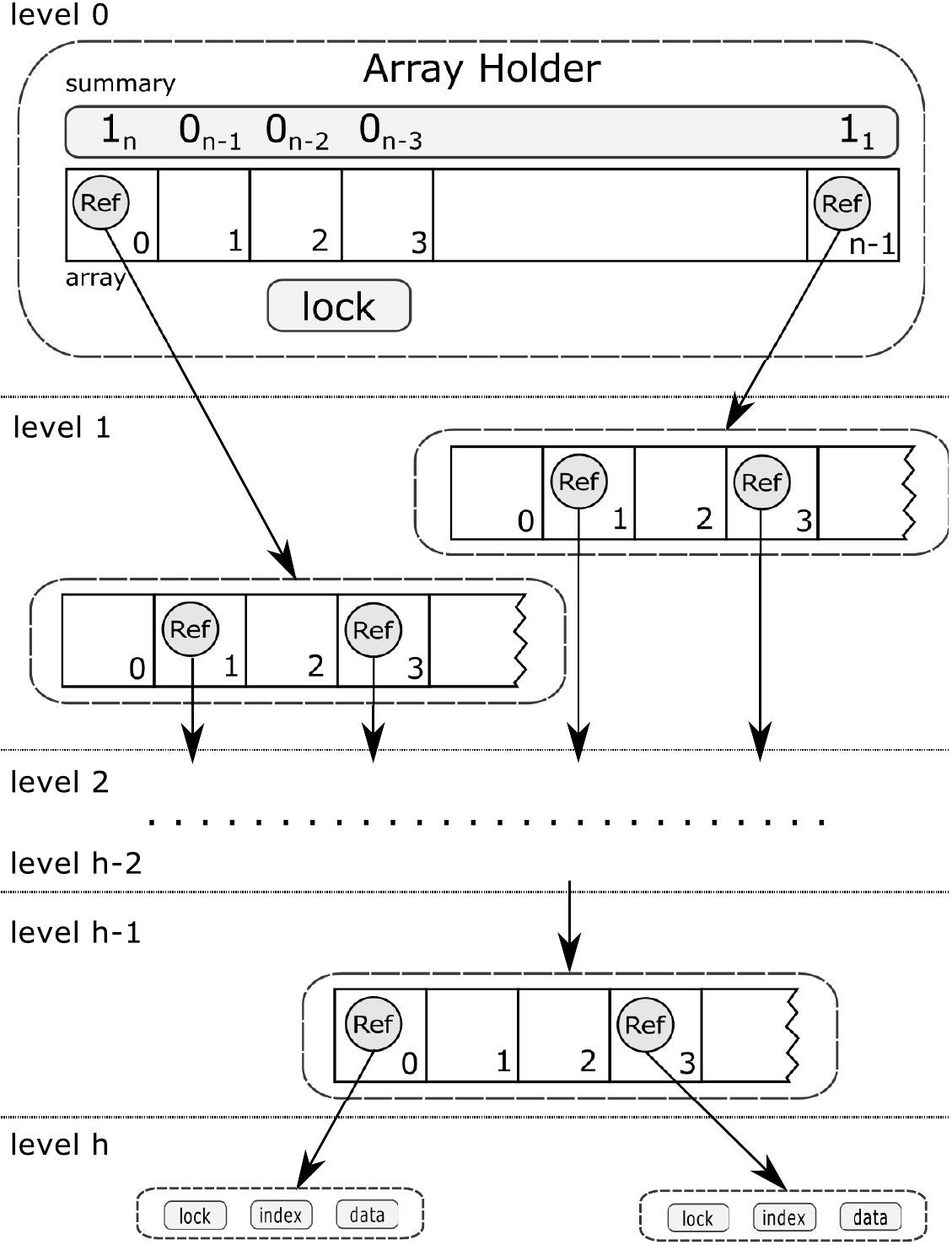}
\par\end{centering}

\caption{\emph{dcvEB array} scheme}
\label{fig:dcvEB-array-scheme}

\end{figure}

\subsection{dcvEB array methods\label{sub:dcvEB-array-methods}}

The \emph{dcvEB} array is designed to support all the dynamic set
methods as specified in \cite[p. 230]{Cormen2009ita}. Not all of
them are extensively discussed in the article, although all of them
are implemented\footnotemark[1]. In particular, the basics of the
missing \emph{predecessor()} method are very similar to \emph{successor(),}
which is discussed below, whilst the methods \emph{minimum()} and
\emph{maximum()} have straightforward implementation using \emph{successor()}
and \emph{predecessor()}\footnote{The minimum can be determined by the call \emph{successor(0)}, whilst
maximum by the call \emph{predecessor(}$n^{h}-1$\emph{)} }. It is assumed that the stored objects are uniquely identified by
integer keys. Thus, the key appears in most of the \emph{dcvEB }array
methods as an input parameter, whilst the return value of all the
query methods is the pair consisting of the key and the stored element.
The presented implementation uses locks as well as lock-free synchronization
mechanisms. Hence, wherever an atomic, lock-free element is used,
an appropriate object or variable is declared as atomic. The main
purpose of this section is to allow the reader to understand the general
idea behind the presented algorithms and the data structures they
use. For this reason some issues connected with synchronization and
concurrency are only indicated, and will be discussed later. 

The methods presented above use two additional structures: \emph{ArrayHolder}
(Listing: \ref{listing:ArrayHolderStructure}), \emph{ArrayParam},
and one atomic common variable \emph{ap}, which holds the current
\emph{ArrayParam} value. The \emph{ArrayHolder} contains five fields:
\emph{array} - atomic array of references to the lower-level \emph{AHs},
summary - an atomic bit vector implemented as any integer type available
on the current hardware platform, \emph{index} - an atomic key value
of the stored object, \emph{data} - an atomic reference to the stored
object, and \emph{lock} - \emph{reader-writer} lock object associated
with the given $AH$. 

\SetAlFnt{\small\sf}
\IncMargin{2em}
\begin{algorithm}

\texttt{ArrayHolder}

\texttt{~~AtomicRefArray array}

\texttt{~~AtomicInt summary}

\texttt{~~AtomicInt index}

\texttt{~~AtomicRef data}

\texttt{~~RWLock lock}

\NoCaptionOfAlgo
\caption{\textbf{Listing \thealgocf:} Array Holder structure} \label{listing:ArrayHolderStructure}

\end{algorithm}
\DecMargin{2em}

The second structure \emph{ArrayParam} (\emph{AP}) contains the fields:
\emph{size} - the number of indices assignable at the moment in the
\emph{dcvEB} array (i.e. the maximal object stored in the \emph{dcvEB}
array cannot have a key greater than \emph{$size-1$}), \emph{height}
- the number of levels of a tree implementing the \emph{dcvEB} array
structure except the last leaf level, and \emph{root} - a root's $AH$
(Listing: \ref{listing:ArrayParam}).

\SetAlFnt{\small\sf}
\IncMargin{2em}
\begin{algorithm}

\texttt{ArrayParam}

\texttt{~~int size}

\texttt{~~int height}

\texttt{~~ArrayHolder root}

\NoCaptionOfAlgo
\caption{\textbf{Listing \thealgocf:} Array Parameters structure} \label{listing:ArrayParam}

\end{algorithm}
\DecMargin{2em}

The current value of \emph{ArrayParam} is stored in the common atomic
variable \emph{ap}. Except for initialization, the fields of AP are
read-only, hence they do not need to be synchronized. 

The first presented method discussed in this section is \emph{insert()}
(Listing: \ref{listing:cvEBarray:Insert}). At the very beginning
it locks the common atomic variable \emph{ap} (in order to prevent
altering the current \emph{ArrayParam} reference by \emph{remove()}),
then it makes the local copy of the current array parameters (Line:
\ref{code:Insert:ap2cap}). Next, it locks the root, (Line: \ref{code:Insert:root:rLock}),
and unlocks \emph{ap} (Line: \ref{code:Insert:ap:rUnlock}). 

Then, it checks whether the key value fits the current array size
and, if not, it tries to extend the array (Listing: \ref{listing:cvEBarray:Insert},
Lines: \ref{code:Insert:array-grow} - \ref{code:Insert:setting-new-top}).
Array growing is implemented by adding successive levels above the
current root (Listing: \ref{listing:cvEBarray:arrayGrowing}). When
the new top of the \emph{dcvEB array} tree is ready, the algorithm
tries to set it as the new root within the newly created \emph{ArrayParam}
record (Listing: \ref{listing:cvEBarray:arrayGrowing}, Line: \ref{code:Insert:setting-new-top}).
Then, irrespectively of the result of the \emph{CAS}\footnote{\emph{CAS(a,b,c)} - compare and swap atomic action operating\emph{
}under the scheme: if $a=b$ then $a\leftarrow$ c and return \emph{true}.
Return \emph{false} otherwise. } invoke, it unlocks \emph{cAP's} root (Listing: \ref{listing:cvEBarray:Insert},
Line: \ref{code:Insert:cAP:root:unLock}). If, due to concurrent interference
with other threads, \emph{CAS} fails and the common array parameters
are not changed, the root locking guarded by the \emph{ap} lock is
repeated (Listing: \ref{listing:cvEBarray:Insert}, Lines: \ref{code:Insert:apLock-lock}
- \ref{code:Insert:apLock-unLock}), and the loop condition is re-evaluated
(Listing: \ref{listing:cvEBarray:Insert}, Line: \ref{code:Insert:size-check}).
If \emph{CAS} succeeds, then \emph{cAP} is updated (Line: \ref{code:Insert:cAP-to-newAp}),
and the loop is interrupted. 

After the size of the array has been adapted to the size of a key,
the algorithm traverses the tree structure starting from the current
root (Listing: \ref{listing:cvEBarray:Insert}, Line: \ref{code:Insert:currentAH})
to the leaf. On every step of the loop while (Listing: \ref{listing:cvEBarray:Insert},
Lines: \ref{code:Insert:loop:start} - \ref{code:Insert:loop:stop})
a subsequent level of the tree is visited. The loop starts from calculating
the level position \emph{lp}, then, if it is not the top level, \emph{cAH}
becomes read locked, and the previous node \emph{pAH} is unlocked
(Listing: \ref{listing:cvEBarray:Insert}, Line: \ref{code:Insert:rLock}).
Next, \emph{pAH} is set to \emph{cAH}, and \emph{cAH} is atomically
updated (Listing: \ref{listing:cvEBarray:Insert}, Lines: \ref{code:pah-update}
- \ref{code:Insert:setBit}). This update is to set the \emph{$n-lp$}
bit in \emph{summary} corresponding to the $lp$ cell of the \emph{cAH's}
\emph{array} field. After setting the bit indicating that at $lp$
position in $cAH$'s array there is a subtree, iteration moves to
the lower level of the tree, i.e. the current value of $cAH$ is replaced
by the reference to its $lp$ children (Listing: \ref{listing:cvEBarray:Insert},
Line: \ref{code:Insert:currentAHreset}). 

\SetAlFnt{\small\sf}
\begin{algorithm}

\texttt{insert(key, data)}

\texttt{~~apLock.rLock();\label{code:Insert:ap:rLock} cAP $\leftarrow$
ap;\label{code:Insert:ap2cap}}

\texttt{~~cAP.root.rLock();\label{code:Insert:root:rLock} apLock.rUnlock();\label{code:Insert:ap:rUnlock}}

\texttt{~~while (key >= cAP.size)\label{code:Insert:size-check}\label{code:Insert:while--growing-start}}

\texttt{~~~~ArrayParam newAP $\leftarrow$ grow(key);\label{code:Insert:array-grow}}

\texttt{~~~~newAP.rLock();}

\texttt{~~~~tmp $\leftarrow$ CAS(ap,cAP,newAP);\label{code:Insert:setting-new-top}}

\texttt{~~~~cAP.root.rUnlock();\label{code:Insert:cAP:root:unLock}}

\texttt{~~~~if not tmp then}

\texttt{~~~~~~apLock.rLock();\label{code:Insert:apLock-lock}
cAP $\leftarrow$ ap;\label{code:Insert:cAP-update}}

\texttt{~~~~~~cAP.root.rLock();apLock.rUnlock();\label{code:Insert:apLock-unLock}}

\texttt{~~~~else cAP $\leftarrow$ newAp;\label{code:Insert:cAP-to-newAp}
break;}

\texttt{~~end while;\label{code:Insert:while-stop}}

\texttt{~~cAH $\leftarrow$ cAP.root;\label{code:Insert:currentAH}
cl $\leftarrow$ $0$;\label{code:Insert:currentLevel} pAH $\leftarrow$
nil;\label{code:Insert:previousAH:nil}}

\texttt{~~while (cl $<$ cAP.height)\label{code:Insert:loop:start}}

\texttt{~~~~lp $\leftarrow$ lvlPos(cl,~key);\label{code:Insert:computeLevelPos}}

\texttt{~~}\texttt{\textbf{~~}}\texttt{if cl $\neq$ $0$ then}

\texttt{~~~~~~cAH.rLock();\label{code:Insert:rLock} pAH.rUnlock();\label{code:Insert:rUnlock1}}

\texttt{~~}\texttt{\textbf{~~}}\texttt{pAH $\leftarrow$ cAH;\label{code:pah-update}
cAH.summary~$\leftarrow$ \label{code:summary}}

\texttt{~~~~~~$0_{n}\ldots0_{n-lp+1}1_{n-lp}0_{n-lp-1}\ldots0_{1}$}

\texttt{~~~~~~$OR_{bit}$~cAH.summary;\label{code:Insert:setBit}}

\texttt{~~~~cAH $\leftarrow$ cAH.array{[}lp{]};\label{code:Insert:currentAHreset}}

\texttt{~~~~if (cAH = nil) then\label{code:Insert:subtree-initialization-req}}

\texttt{~~~~~~cAH~$\leftarrow$~createAH();\label{code:Insert:subtree-AH-creation}}

\texttt{~~~~~~CAS(pAH.array{[}lp{]},nil,cAH)\label{code:Insert:assign-to-a-parent-array}}

\texttt{~~~~~~cAH $\leftarrow$ pAH.array{[}lp{]};\label{code:Insert:cAHfinalValue}}

\texttt{~~}\texttt{\textbf{~~}}\texttt{cl $\leftarrow$ cl+1;\label{code:Insert:currentLevelInc}}

\texttt{~~end while \label{code:Insert:loop:stop}}

\texttt{~~cAH.data $\leftarrow$ data;\label{code:Insert:dataSet}
cAH.index $\leftarrow$ key;\label{code:Insert:keySet}}

\texttt{~~pAH.rUnlock();\label{code:Insert:previousAH:unlock}}

\NoCaptionOfAlgo
\caption{\textbf{Listing \thealgocf:} Insert method} \label{listing:cvEBarray:Insert}

\end{algorithm} 

Of course, it is possible that the subtree has not yet been initialized
(Listing: \ref{listing:cvEBarray:Insert}, Line: \ref{code:Insert:subtree-initialization-req}).
In such a case the new $cAH$ is created, atomically assigned to the
parent $AH$'s array when possible (Listing: \ref{listing:cvEBarray:Insert},
Line: \ref{code:Insert:assign-to-a-parent-array}), then due to the
possible interference with another insert thread (but not remove thread)
the final value of $cAH$ is re-read from the parent $cAH$'s array
(Listing: \ref{listing:cvEBarray:Insert}, Line: \ref{code:Insert:cAHfinalValue}).
At the end of the loop, the variable determining the current level
of iteration is incremented (Listing: \ref{listing:cvEBarray:Insert},
Line: \ref{code:Insert:currentLevelInc}). The loop ends when $cAP$
is pointing at some leaf $AH$. Hence, at the end of the method both
leaf $AH$'s fields: data and index, are updated. In the last line
of \emph{insert()} the leaf's parent node lock is released (Listing:
\ref{listing:cvEBarray:Insert}, Line: \ref{code:Insert:previousAH:unlock}).

An important routine used within the \emph{insert()} method is \emph{grow()}.
It is responsible for extending the \emph{dcvEB array}, when it is
too small to hold an element with the given \emph{key}. Enlarging
the array relies on adding additional levels above the existing root
so that the total height $h$ of the \emph{dcvEB array} tree increases.
Hence, the \emph{dcvEB array} becomes capacious enough to encompass
the\emph{ key }i.e. it requires $n^{h}>$\emph{~key.} As a result
of this operation, a new \emph{AP} record is created (Listing: \ref{listing:cvEBarray:arrayGrowing},
Line: \ref{code:grow:createNewAP}). Then, the \emph{grow()} procedure
calculates the appropriate new height and size (Listing: \ref{listing:cvEBarray:arrayGrowing},
Lines: \ref{code:grow:newHight} - \ref{code:grow:newSize}). The
number of levels to create is determined as the difference between
the previous height and the new height of the tree (Listing: \ref{listing:cvEBarray:arrayGrowing},
\ref{code:Grow:TopSize}). Then, the procedure starts the loop \emph{while}
(Listing: \ref{listing:cvEBarray:arrayGrowing}, Lines: \ref{code:Grow:start-while}
- \ref{code:Grow:end-while}), and within every turn of the loop the
new $AH$ is generated. The first generated $AH$ becomes a new root
of the tree (Listing: \ref{listing:cvEBarray:arrayGrowing}, Line:
\ref{code:Grow:new-root-assignment}), each further one becomes the
leftmost child of its predecessor (Listing: \ref{listing:cvEBarray:arrayGrowing},
Line: \ref{code:Grow:AH-chaining}), and finally the last generated
$AH$ takes the previous root $AH$ as its leftmost child (Listing:
\ref{listing:cvEBarray:arrayGrowing}, Line: \ref{code:Grow:adopt-the-previous-root}).
The sequential running time of \emph{grow()} is $O(\log_{n}\alpha)$.
Since the number of iterations of the loop while (Listing: \ref{listing:cvEBarray:Insert},
Lines: \ref{code:Insert:while--growing-start} - \ref{code:Insert:while-stop})
depends on the interferences with the concurrently operating delete
threads, whilst the number of iterations of \emph{while} (Listing:
\ref{listing:cvEBarray:Insert}, Lines: \ref{code:Insert:loop:start}
- \ref{code:Insert:previousAH:unlock}) is limited by the height $h=\log_{n}\alpha$
of the tree, then the overall sequential running time of \emph{insert()}
is $O(\log_{n}\alpha)$.

\SetAlFnt{\small\sf}
\begin{algorithm}

\texttt{grow(key)}

\texttt{~~nAP $\leftarrow$ createAP()\label{code:grow:createNewAP}}

\texttt{~~nAP.height $\leftarrow$ $\left\lceil \log_{d}\right.$key$\left.\right\rceil $\label{code:grow:newHight}}

\texttt{~~nAP.size $\leftarrow$ $d^{\mbox{cAP.hight}}$\label{code:grow:newSize}}

\texttt{~~cAP $\leftarrow$ ap;\label{code:grow:fetchAP}}

\texttt{~~topSize $\leftarrow$ nAP.height - cAP.height\label{code:Grow:TopSize}}

\texttt{~~cl $\leftarrow$ $0$}

\texttt{~~while (cl $<$ topSize)\label{code:Grow:start-while} }

\texttt{~~~~cAH $\leftarrow$ createAH()}

\texttt{~~~~cAH.summary $\leftarrow$ $1_{n}0_{n-1}\ldots0_{1}$}

\texttt{~~~~if cl = 0 }

\texttt{~~~~~~nAP.root $\leftarrow$ cAH\label{code:Grow:new-root-assignment}}

\texttt{~~~~else}

\texttt{~~~~~~pAH.array{[}0{]} $\leftarrow$ cAH \label{code:Grow:AH-chaining}}

\texttt{~~~~if cl = topSize - $1$ }

\texttt{~~~~~~cAH.array{[}0{]} $\leftarrow$ cAP.root \label{code:Grow:adopt-the-previous-root}}

\texttt{~~~~pAH $\leftarrow$ cAH\label{code:Grow:end-while}}

\texttt{~~return nAP}

\NoCaptionOfAlgo
\caption{\textbf{Listing \thealgocf:} Array growing} \label{listing:cvEBarray:arrayGrowing}

\end{algorithm}

The next method \emph{get()}, similarly to \emph{insert()}, first
retrieves the current snapshot of the \emph{dcvEB array} parameters
(Listing: \ref{listing:dcvEBarray:Get}, Line: \ref{code:get:ap2cap}),
then traverses the structure down from the root to the leaf following
the subsequent level positions. The main difference between \emph{get()}
in the \emph{dcvEB} array and \emph{get()} from the previous version
of the structure \cite{Kulakowski2014acve} is that currently the
enabled bit in a summary does not guarantee the existence of the corresponding
lower level array holder. Hence, the additional check whether the
next \emph{AH} is not actually \emph{nil} is necessary (Listing: \ref{listing:dcvEBarray:Get},
Lines: \ref{code:Get:cAH:lowlevcheck-start} - \ref{code:Get:cAH:lowlevcheck-end}).
As can be seen, the sequential running time of \emph{get()} is determined
by the loop (Listing: \ref{listing:dcvEBarray:Get}, Lines: \ref{code:Get:while:start}
- \ref{code:Get:while:end}) and is $O(\log_{n}\alpha)$. 

\SetAlFnt{\small\sf}
\begin{algorithm}

\texttt{get(key)}

\texttt{~~cAP $\leftarrow$ ap;\label{code:get:ap2cap} cAH $\leftarrow$
cAP.root; cl $\leftarrow$ $0$;}

\texttt{~~while (cl $<$ cAP.height)\label{code:Get:while:start}}

\texttt{~~~~lp~$\leftarrow$ lvlPos(cl,~key)}

\texttt{~~}\texttt{\textbf{~~}}\texttt{if~($0_{n}0_{n-1}\ldots0_{n-lp+1}1_{n-lp}0_{n-lp-1}\ldots0_{1}$}

\texttt{~~~~~~~$AND_{bit}$~cAH.summary)~=~$0$\label{code:Get:getBit}}

\texttt{~~}\texttt{\textbf{~~}}\texttt{~~return nil\label{code:Get:GetBit:returnNil}}

\texttt{~~~~cAH $\leftarrow$ cAH.array{[}lp{]}}

\texttt{~~~~if cAH = nil \label{code:Get:cAH:lowlevcheck-start}}

\texttt{~~~~~~return nil\label{code:Get:cAH:lowlevcheck-end}}

\texttt{~~}\texttt{\textbf{~~}}\texttt{cl++}

\texttt{~~end while\label{code:Get:while:end}}

\texttt{~~return (cAH.data, cAH.index)\label{code:Get:ReturnResult}}

\NoCaptionOfAlgo
\caption{\textbf{Listing \thealgocf:} Get method} \label{listing:dcvEBarray:Get}

\end{algorithm}

Changes resulting from the introduction of dynamic memory allocation
also affected the \emph{delete()} method. Since \emph{insert()} is
able to expand the top of the \emph{dcvEB array} tree and to generate
missing lower level \emph{AHs}, then \emph{delete()} needs to be able
to trim the top of the tree and to remove redundant nodes. The \emph{delete()}
method implementation can be logically divided into three stages:
preparing a path towards a leaf, deleting the leaf with the deletion
propagation and cleaning, and the \emph{dcvEB} array top trimming.
Like almost all presented dynamic set methods, \emph{delete()} also
starts from fetching the snapshot of the current \emph{dcvEB} array
parameters (Listing: \ref{listing:dcvEBarray:Delete}, Lines: \ref{code:delete:ap2cap}
- \ref{code:delete:curLevSet}). Then, after the creation of the two
empty tables \emph{ahol} and \emph{pos} for holding the path between
the root and the node for disposal, the method \emph{makePath} is
invoked (Listing: \ref{listing:dcvEBarray:Delete}, Line: \ref{code:Delete:PreparePathToLeafCall}).
The purpose of this method is to fill these tables with the subsequent
\emph{AH}s and their positions along the way from the root to the
leaf node being removed according to the formula for $lp_{k}$. During
the iteration, similarly to in \emph{get()}, the presence of the child
must be checked twice. Firstly, by checking a~summary bit vector
(Listing: \ref{listing:cvEBarray:PreparePathToLeaf}, Lines: \ref{code:PreparePath:summary:check:begin}
- \ref{code:PreparePath:summary:check:end}), the second time by checking
whether the retrieved subsequent \emph{AH} is not \emph{nil} (Listing:
\ref{listing:cvEBarray:PreparePathToLeaf}, Line: \ref{code:PreparePath:summary:check2:begin}).
It is assumed that the arguments of \emph{makePath} are in-out, which
means that the changes made inside the method are visible outside. 

\SetAlFnt{\small\sf}
\begin{algorithm}

\texttt{delete(key)}

\texttt{~~cAP $\leftarrow$ ap;\label{code:delete:ap2cap} cAH $\leftarrow$
cAP.root; }

\texttt{~~pAH $\leftarrow$ nil; cl $\leftarrow$ $0$;\label{code:delete:curLevSet}}

\texttt{~~ahol $\leftarrow$ makeEmptyArray(cAP.height)}

\texttt{~~pos $\leftarrow$ makeEmptyArray(cAP.height)\label{code:Delete:InitBlock:position}}

\texttt{~~makePath(key,cAP,cAH,pAH,cl,ahol,pos)\label{code:Delete:PreparePathToLeafCall}}

\texttt{~~if} (\texttt{!is\_filed(ahol,cAP.height)) then\label{code:Delete:fill:chcek}}

\texttt{~~~~return;\label{code:Delete:fill:chcek-over}}

\texttt{~~delIntern(key,cAP,cAH,pAH,cl,ahol,pos);}

\texttt{~~while (cAP != ap and rep < maxRep)\label{code:Delete:while-with-deleteClean-begin}}

\texttt{~~~~deleteClean(key);\label{code:Delete:deleteClean}\label{code:Delete:while-with-deleteClean-end}
rep $\leftarrow$ rep + 1;}

\texttt{~~topTrim();}

\NoCaptionOfAlgo
\caption{\textbf{Listing \thealgocf:} Delete method} \label{listing:dcvEBarray:Delete}

\end{algorithm}

It is noteworthy that \emph{makePath} may not contain a complete path
between the root and the leaf designed to be disposed. This happens
when there is no such path i.e. because the desired element has just
been removed. In such a case the procedure stops, and leaves the arrays
\emph{ahol} and \emph{pos} partially filled. In the case of \emph{delete()},
not entirely filled arrays indicate that there is no element to delete
(Listing: \ref{listing:dcvEBarray:Delete}, Line: \ref{code:Delete:fill:chcek}).
Hence the method can finalize its operation (Listing: \ref{listing:dcvEBarray:Delete},
Line: \ref{code:Delete:fill:chcek-over}).

\SetAlFnt{\small\sf}
\begin{algorithm}

\texttt{makePath(key,cAP,cAH,pAH,cl,ahol,pos)}

\texttt{~~while (cl $<$ cAP.height)\label{code:PreparePath:while1:begin}}

\texttt{~~~~lp $\leftarrow$ lvlPos(cl,~key);}

\texttt{~~~~if~($0_{n}\ldots0_{n-lp+1}1_{n-lp}0_{n-lp-1}\ldots0_{1}\,\,AND_{bit}$\label{code:PreparePath:summary:check:begin}}

\texttt{~~~~~~cAH.summary)~=~$0$ then break;\label{code:PreparePath:summary:check:end}}

\texttt{~~~~pos{[}cl{]} $\leftarrow$ lp; ahol{[}cl{]} $\leftarrow$
cAH;}

\texttt{~~~~pAH $\leftarrow$ cAH; cAH $\leftarrow$ cAH.array{[}lp{]};}

\texttt{~~~~if (cAH = nil) then\label{code:PreparePath:summary:check2:begin}
break;\label{code:PreparePath:summary:check2:end}}

\texttt{~~}\texttt{\textbf{~~}}\texttt{cl $\leftarrow$ cl~+~$1$;}

\texttt{~~end while\label{code:PreparePath:while1:end}}

\NoCaptionOfAlgo
\caption{\textbf{Listing \thealgocf:} makePath - delete auxiliary method}\label{listing:cvEBarray:PreparePathToLeaf}

\end{algorithm}

The second and the major subroutine of \emph{delete()} is \emph{delIntern()}.
In terms of the synchronization structure, it is similar to the original
\emph{delete()} method presented in \cite{Kulakowski2014acve}. The
need, however, for effective array holder removal caused the necessity
to introduce a few new elements into the code of the algorithm. The
\emph{delIntern()} method is executed only if the array holder structure
is correctly filled, which takes place only if \emph{makePath()} (Listing:
\ref{listing:cvEBarray:PreparePathToLeaf}) does not break its while
loop. Hence, at the very beginning of \emph{delIntern()}, it is assumed
that the variable \emph{pAH} refers to some \emph{AH} from the last
but one (\emph{cAP.height-1}) level, whilst \emph{cAH} points at the
element from the last level containing pairs \emph{(index, data)}. 

Thus, after locking appropriate array holders (Listing: \ref{listing:dcvEBarray:internalDelete},
Line: \ref{code:internalDelete:initialLocking}), the stored data
are overwritten by \emph{nil} (Listing: \ref{listing:dcvEBarray:internalDelete},
Line: \ref{code:internalDelete:dataOverwritten}). Afterwords \emph{delIntern()}
begins its arduous journey towards the root iterating within the loop
while (Listing: \ref{listing:dcvEBarray:internalDelete}, Lines: \ref{code:Delete:while2:begin}
- \ref{code:Delete:while2:end}). It starts from the last but one
level (Listing: \ref{listing:dcvEBarray:internalDelete}, Line:\texttt{ }\ref{code:Delete:while2:last-but-one-level-if}).
First, it sets an appropriate bit in \emph{cAH's} summary to $0$
(Listing: \ref{listing:dcvEBarray:internalDelete}, Line:\texttt{ }\ref{code:Delete:while2:bit-set}).
Therefore, the data was logically removed from the structure (data
field is set to nil, index to $-1$, and AH is not by the parent's
summary), although an appropriate array holder still exists. Such
an array holder will be physically removed only when the whole \emph{cAH's}
\emph{summary} is $0$ (Listing: \ref{listing:dcvEBarray:internalDelete},
Line: \ref{code:Delete:physically-remove}). Otherwise, if \emph{cAH's
summary} is not $0$, the previously locked nodes are released and
the method exits (Listing: \ref{listing:dcvEBarray:internalDelete},
Lines: \ref{code:Delete:while2:cAHsummaryNEQ0} - \ref{code:Delete:while2:cAHsummaryNEQ0-end}). 

\SetAlFnt{\small\sf}
\begin{algorithm}

\texttt{delIntern(key,cAP,cAH,pAH,cl,ahol,pos)}

\texttt{~~pAH.wLock(); cAH.wLock(); \label{code:internalDelete:initialLocking}}

\texttt{~~cAH.data $\leftarrow$ nil;\label{code:internalDelete:dataOverwritten}
cAH.index $\leftarrow-1$; }

\texttt{~~pAH $\leftarrow$ cAH;}

\texttt{~~while (cl $\geq$ $0$)\label{code:Delete:while2:begin}}

\texttt{~~~~cAH $\leftarrow$ ahol{[}cl{]};\label{code:Delete:while2:parentAH}
lp $\leftarrow$ pos{[}cl{]};}

\texttt{~~~~if (cl = cAP.height - 1)\label{code:Delete:while2:last-but-one-level-if}}

\texttt{~~~~~~cAH.summary $\leftarrow$ $1_{n}\ldots0_{n-lp}\ldots1_{1}$}

\texttt{~~~~~~~$AND_{bit}$~cAH.summary; \label{code:Delete:while2:bit-set}}

\texttt{~~~~~~if (cAH.summary $\neq$ 0) \label{code:Delete:while2:cAHsummaryNEQ0} }

\texttt{~~~~~~~~pAH.wUnlock(); \label{code:Delete:while2:linearization_point_1}
cAH.wUnlock(); }

\texttt{~~~~~~~~return;\label{code:Delete:while2:cAHsummaryNEQ0-end}}

\texttt{~~~~~~else}

\texttt{~~~~~~~~cAH.array $\leftarrow$ $\{nil_{0},\ldots,nil_{n-1}\}$;
\label{code:Delete:physically-remove}}

\texttt{~~~~else}

\texttt{~~~~~~cAH.wLock(); pAH.wLock();}

\texttt{~~~~~~isSummaryAltered~$\leftarrow$~false }

\texttt{~~~~~~if (pAH.summary = $0$)\label{code:Delete:while2:bitVectorCheck}}

\texttt{~~~~~~~~cAH.summary~$\leftarrow$~$1_{n}\ldots0_{n-lp}\ldots1_{1}$\label{parentAH.summary.bitresetting-start-1}}

\texttt{~~~~~~~~~$AND_{bit}$~pAH.summary;\label{parentAH.summary.bitresetting-end-1}}

\texttt{~~~~~~~~isSummaryAltered~$\leftarrow$~true }

\texttt{~~~~~~if (cAH.summary = $0$)\label{parentAH.summary.summary-check-begin}}

\texttt{~~~~~~~~cAH.array $\leftarrow$ $\{nil_{0},\ldots,nil_{n-1}\}$\label{parentAH.summary.summary-check-end}}

\texttt{~~~~~~cAH.wUnlock();\label{code:Delete:cAH.wunlock}
pAH.wUnlock();\label{code:Delete:pAH.wunlock}\label{code:Delete:while2:linearization_point_2}}

\texttt{~~~~~~if (not isSummaryAltered)\label{code:Delete:while2:parentCheck:start}}

\texttt{~~~~~~~~return;\label{code:Delete:break-propagation}}

\texttt{~~~~cl $\leftarrow$ cl~-~$1$;\label{code:Delete:cl-update}
pAH $\leftarrow$ cAH;\label{code:Delete:pAH-update}}

\texttt{~~end while\label{code:Delete:while2:end}}

\NoCaptionOfAlgo
\caption{\textbf{Listing \thealgocf:} Delete internal - delete auxiliary method} \label{listing:dcvEBarray:internalDelete}

\end{algorithm}

This, \textquotedbl{}my brother keeps me alive\textquotedbl{}, lazy
strategy aims to reduce the amount of memory allocation performed
during the course of the algorithm. If \emph{delIntern()} processes
the element on the level \emph{cAP.height $-2$} or higher, then it
first locks the current and previous \emph{AH's} node, and next \emph{alters}
the \emph{cAH} summary by removing the bit corresponding to the removed
children (Listing: \ref{listing:dcvEBarray:internalDelete}, Line:
\ref{parentAH.summary.bitresetting-end-1}). As in the previous case,
if \emph{cAH} summary is $0$ then the child node is dereferenced
(Listing: \ref{listing:dcvEBarray:internalDelete}, Lines: \ref{parentAH.summary.summary-check-begin}
- \ref{parentAH.summary.summary-check-end}). Then, after unlocking
\emph{cAH} and \emph{pAH} (Listing: \ref{listing:dcvEBarray:internalDelete},
Line: \ref{code:Delete:cAH.wunlock}) and checking whether it makes
sense to propagate a delete action towards the root (Listing: \ref{listing:dcvEBarray:internalDelete},
Lines: \ref{code:Delete:while2:parentCheck:start} - \ref{code:Delete:break-propagation})
the \emph{else} block ends. At the end of the procedure the variables
\emph{cl} (current level) and \emph{pAH} (previous array holder) are
updated (Listing: \ref{listing:dcvEBarray:internalDelete}, Line:
\ref{code:Delete:cl-update}). 

The next subroutine of \emph{delete()} is \emph{deleteClean()}. It
is called from \emph{delete()} just after \emph{delIntern()} (Listing:
\ref{listing:dcvEBarray:Delete}, Lines: \ref{code:Delete:while-with-deleteClean-begin}
- \ref{code:Delete:while-with-deleteClean-end}). The main reason
for which it is introduced is the danger of not removing all the required
\emph{AH} when the delete action interferes with the insert action.
The idea of \emph{deleteClean()} implementation and further explanations
are in Subsection \ref{sub:Expanding-and-trimming}.

\SetAlFnt{\small\sf}
\begin{algorithm}

\texttt{topTrim()}

\texttt{~~cAP $\leftarrow$ ap;}

\texttt{~~while (cAP.root.summary = $1_{n}0_{n-1}\ldots0_{1}$)\label{code:toptrim:while:begin}}

\texttt{~~~~if (cAP.height = 1)\label{code:toptrim:trim-stop}
then break;\label{code:toptrim:trim-stop-break} }

\texttt{~~~~nAP $\leftarrow$ createAP();\label{code:toptrim:nAP-creation}}

\texttt{~~~~nAP.height $\leftarrow$ cAP.height - 1;}

\texttt{~~~~nAP.size $\leftarrow$ $d^{\mbox{nAP.hight}}$;\label{code:toptrim:nAP-initialization-end}}

\texttt{~~~~theLonelyChild $\leftarrow$ }

\texttt{~~~~~~cAP.root.array{[}$0${]};\label{code:toptrim:lonelyChild-retrieving}}

\texttt{~~~~if (theLonelyChild = nil) then }

\texttt{~~~~~~return nil;}

\texttt{~~~~nAP.root $\leftarrow$ theLonelyChild;\label{code:toptrim:lonelyChild-to-root-candidate-promotion}}

\texttt{~~~~apLock.wLock();\label{code:toptrim:apLock} cAP.root.wLock();\label{code:toptrim:root:wLock}}

\texttt{~~~~if (cAP.root.summary = $1_{n}0_{n-1}\ldots0_{1}$)\label{code:toptrim:if:trimming-needed}}

\texttt{~~~~~~then CAS(ap,cAP,nAP);\label{code:toptrim:top:update}}

\texttt{~~~~cAP.root.wUnlock();\label{code:toptrim:root:wUnLock}apLock.wUnlock();\label{code:toptrim:apUnLock}}

\texttt{~~~~cAP $\leftarrow$ ap;\label{code:toptrim:while:end}}

\NoCaptionOfAlgo
\caption{\textbf{Listing \thealgocf:} TopTrim - delete auxiliary method} \label{listing:cvEBarray:topTrim}

\end{algorithm}

The purpose of \emph{topTrim()} - the last auxiliary method involved
in \emph{delete()} implementation is to cut the top of the \emph{dcvEB}
array tree if it is reduced to the list (Listing: \ref{listing:cvEBarray:topTrim}).
It is possible that, after the \emph{internalDelete()} call, the root
and a few nodes below have only one, the leftmost, child. In such
a case, the sequence of such vertices starting from the root needs
to be safely removed. The \emph{topTrim()} reduces the top of the
tree iteratively. It removes only one node (root) in every course
of the loop while (Listing: \ref{listing:cvEBarray:topTrim}, Lines:
\ref{code:toptrim:while:begin} - \ref{code:toptrim:while:end}).
If the loop \emph{while} condition is met, i.e. the root has only
one child at the leftmost cell in the array, then the new \emph{AP}
candidate is prepared (Listing: \ref{listing:cvEBarray:topTrim},
Lines: \ref{code:toptrim:nAP-creation} - \ref{code:toptrim:nAP-initialization-end}).
Next the ``lonely'' child is retrieved (Listing: \ref{listing:cvEBarray:topTrim},
Line: \ref{code:toptrim:lonelyChild-retrieving}). If it is not nil
(it might be nil due to another delete thread), it is promoted to
a new root candidate of the whole \emph{dcvEB} array tree (Listing:
\ref{listing:cvEBarray:topTrim}, Line: \ref{code:toptrim:lonelyChild-to-root-candidate-promotion}).
Finally, if the array properties are not changed during the course
of the \emph{topTrim} routine (i.e. the assertion that the root has
only one leftmost child still holds) the newly prepared \emph{nAP}
becomes the main array parameters reference. The \emph{topTrim()}
method does not trim the trees shallower than the ones composed of
the root and leaves (Listing: \ref{listing:cvEBarray:topTrim}, Line:
\ref{code:toptrim:trim-stop}). The \emph{CAS} call (Listing: \ref{listing:cvEBarray:topTrim},
Line: \ref{code:toptrim:top:update}) responsible for the \emph{ArrayParam}
altering is guarded by two locks (Listing: \ref{listing:cvEBarray:topTrim},
Line: \ref{code:toptrim:apLock}). They prevent a situation in which
\emph{insert()} adds the new element into the subtree rooted in the
node, which is subject to removal by\emph{ topTrim()}. The sequential
running time of \emph{delete()} depends on the complexity of their
subroutines. The first of them \emph{makePath()} (Listing: \ref{listing:cvEBarray:PreparePathToLeaf})
comprises one loop \emph{while}. Due to the loop condition (Listing:
\ref{listing:cvEBarray:PreparePathToLeaf}, Line: \ref{code:PreparePath:while1:begin})
it is clear that the sequential running time of \emph{makePath()}
is limited by the height of the tree i.e. $O(\log_{n}\alpha)$. The
methods \emph{delIntern(),} \emph{deleteClean()} and \emph{topTrim()}
also need at most to visit all the nodes on a single path between
the root and a leaf. Therefore, their sequential running time is $O(\log_{n}\alpha)$.
Furthermore, if only one thread is up and running, the loop while
(Listing: \ref{listing:dcvEBarray:Delete}, Line: \ref{code:Delete:while-with-deleteClean-begin})
executes only one. Thus the overall sequential running time of \emph{delete()}
equals the maximum of the running time of all their subcomponents,
and is $O(\log_{n}\alpha)$.

The pseudo code of the last method \emph{successor()} was divided
into two parts. The first (Listing: \ref{listing:dcvEBarray:successor-part1})
one is responsible for the attempt to reach the leaf \emph{AH} holding
the data indexed by a \emph{key}. If such a leaf exists, it will be
returned as its own successor. The second part (Listing: \ref{listing:dcvEBarray:successor-part2})
contains a loop which consists of two other loops, where the first
internal loop is responsible for traversing the \emph{dcvEB array}
tree up, whilst the second traverses the tree down. Such a structure
of the code in the second part corresponds to the \emph{successor()}'s
searching strategy. In other words, first the method tries to go a
little bit higher to check where a successor leaf could be (the first
internal loop), then tries to go towards the leaf in order to retrieve
the stored data and key (the second internal loop). Of course, sometimes
during the gliding down the tree the successor candidate might be
removed. In such a case, the second loop must be aborted and the method
once again starts to follow up the tree in order to find another potential
successor candidate. 

\SetAlFnt{\small\sf}
\begin{algorithm}

\texttt{successor(key)}

\texttt{~~cAP $\leftarrow$ ap;\label{code:successor:ap-fetch}
cAH $\leftarrow$ cAP.root;\label{code:successor:cAH-preparation}}

\texttt{~~if (cAH.summary$\,$=$\,$0) then return nil;}

\texttt{~~pAH $\leftarrow$ nil;\label{code:successor:pAH-init}
cl $\leftarrow$ $0$;\label{code:successor:curLevSet}}

\texttt{~~ahol $\leftarrow$ makeEmptyArray(cAP.height);}

\texttt{~~pos $\leftarrow$ makeEmptyArray(cAP.height);\label{code:successor:InitBlock:position}}

\texttt{~~makePath(key,cAP,cAH,pAH,cl,ahol,pos);\label{code:successor:PreparePathToLeafCall}}

\texttt{~~if (cl = cAP.height) then\label{code:successor:found-at-once-start}}

\texttt{~~~~cl~$\leftarrow$~cl - 1;}

\texttt{~~~~if (cAH $\neq$ nil) then}

\texttt{~~~~~~data$\,$$\leftarrow$$\,$cAH.data;~index$\,$$\leftarrow$$\,$cAH.index;}

\texttt{~~~~~~if (data $\neq$ nil and index $\neq$ -1)}

\texttt{~~~~~~~~then return pair;\label{code:successor:found-at-once-end}}

\NoCaptionOfAlgo
\caption{\textbf{Listing \thealgocf:} Successor method (part 1)} \label{listing:dcvEBarray:successor-part1}

\end{algorithm}

At the very beginning, the \emph{successor()} sets its own local copy
of the array parameters (Listing: \ref{listing:dcvEBarray:successor-part1},
Line: \ref{code:successor:ap-fetch}), then it prepares a pair of\emph{
}holders, \emph{cAH} and \emph{pAH,} used to traverse the structure.
Then the $cl$ variable indicating the visited level and two other
variables, \emph{ahol} and \emph{pos}, referring to arrays holding
\emph{AH}s and their level positions along the path from the root
to the visited node, are defined (Listing: \ref{listing:dcvEBarray:successor-part1},
Lines: \ref{code:successor:cAH-preparation} - \ref{code:successor:InitBlock:position}).
All the newly introduced variables, including \emph{cAH}, \emph{pAH},
\emph{cl}, \emph{ahol} and \emph{pos} are initiated within the \emph{makePath()}
auxiliary method. If \emph{makePath()} reaches the leaf level, \emph{(}the
condition \emph{cl = cAP.height} is true, see Fig. \ref{fig:dcvEB-array-scheme})
this means that the element indexed by the key exists. Hence, if only
the \emph{successor()} procedure manages to fetch the stored data,
then the appropriate \emph{(data, key)} pair is returned by the method
(Listing: \ref{listing:dcvEBarray:successor-part1}, Lines: \ref{code:successor:found-at-once-start}
- \ref{code:successor:found-at-once-end}). Of course, \emph{makePath()}
may not reach the leaf level (the condition \emph{cl = cAP.height}
does not hold) or even if the leaf is reached, its removal might start
before the leaf data are extracted (one of the following three conditions
is true: \emph{cAH = nil, data = nil, index = -1}). In such a case,
the method control goes to the while loop (Listing: \ref{listing:dcvEBarray:successor-part2})
and the algorithm starts to explore other successor candidates. 

\SetAlFnt{\small\sf}
\begin{algorithm}

\texttt{~~mark: while(true)\label{code:successor:first-outer-loop}}

\texttt{~~~~while (cl $\geq$ 0)\label{code:successor:first-inner-loop-begin}}

\texttt{~~~~~~cAH~$\leftarrow$ ahol{[}cl{]};\label{code:successor:set-cAH}lp~$\leftarrow$
pos{[}cl{]};\label{code:successor:set-lp}}

\texttt{~~~~~~tmpSum = $0_{n}\ldots0_{n-lp}1_{n-lp+1}\ldots1_{1}$
\label{code:successor:tmpSum-set-begin}}

\texttt{~~~~~~~~$AND_{bit}$~cAH.summary;\label{code:successor:tmpSum-set-end}}

\texttt{~~~~~~if (tmpSum = 0) then\label{code:successor:tmpSum-if-cond-start}
cl$\,\leftarrow\,$cl$\,$-$\,$1;\label{code:successor:tmpSum-ifcond-stop}}

\texttt{~~~~~~else break;\label{code:successor:first-inner-loop-break}}

\texttt{~~~~end while\label{code:successor:first-inner-loop-end}}

\texttt{~~~~if (cl = -1) then \label{code:successor:last-level-check}
return nil;\label{code:successor:no-further-successors-nil}}

\texttt{~~~~while(true) \label{code:successor:second-inner-loop-begin}}

\texttt{~~~~~~if (tmpSum = 0) then\label{code:successor:tmpSum-if-zero-restart-start}}

\texttt{~~~~~~~~cl$\,\leftarrow\,$cl$\,$-$\,$1; goto mark;\label{code:successor:tmpSum-if-zero-restart-stop}}

\texttt{~~~~~~bp~$\leftarrow$ mostLeftBitPos(tmpSum);\label{code:successor:most-left-bit}}

\texttt{~~~~~~lp$\,\leftarrow\,$lvlPos(pb);\label{code:successor:lp-set}
pos{[}cl{]}$\,\leftarrow\,$lp;\label{code:successor:pos-set}}

\texttt{~~~~~~ahol{[}cl{]}$\leftarrow$cAH;cAH$\leftarrow$cAH.array{[}lp{]};\label{code:successor:cAH-set}}

\texttt{~~~~~~if (cAH = nil)\label{code:successor:cAH-eq-nil}}

\texttt{~~~~~~~~cl$\,\leftarrow\,$cl$\,$-$\,$1; goto mark;\label{code:successor:goto-mark}}

\texttt{~~~~~~tmpSum~$\leftarrow$ cAH.summary; \label{code:successor:tmpSum-update}cl$\,\leftarrow\,$cl$\,$+$\,$1;}

\texttt{~~~~~~if (cAP.height = cl) then break;\label{code:successor:sec-inn-loop-break}}

\texttt{~~~~end while\label{code:successor:sec-inn-loop-end}}

\texttt{~~~~data$\,\leftarrow\,$cAH.data; index$\,\leftarrow\,$cAH.index;}

\texttt{~~~~if (data$\,\neq\,$nil and index$\,\neq\,$-1) then\label{code:successor:deletion-check}}

\texttt{~~~~~~return (data, index);}

\texttt{~~~~else}

\texttt{~~~~~~cl$\,\leftarrow\,$cl$\,$-$\,$1; goto mark;\label{code:successor:next-iteration-goto}}

\texttt{~~end while;\label{code:successor:main-while-end}}

\NoCaptionOfAlgo
\caption{\textbf{Listing \thealgocf:} Successor method (part 2)} \label{listing:dcvEBarray:successor-part2}

\end{algorithm}

The second part of the \emph{successor()} method (Listing: \ref{listing:dcvEBarray:successor-part2})
is responsible for traversing the structure up and down looking for
the next successor candidate. The first inner loop (Listing: \ref{listing:dcvEBarray:successor-part2},
Lines: \ref{code:successor:first-inner-loop-begin} - \ref{code:successor:first-inner-loop-end})
is responsible for traversing the structure up until the node with
the non-empty subtree further to the right is found or the root level
is achieved, i.e. \emph{cl = 0}. At the very beginning, it initiates
\emph{cAH} and \emph{lp} using the values stored in \emph{pos} and
\emph{ahol} (Listing: \ref{listing:dcvEBarray:successor-part2}, Line:
\ref{code:successor:set-cAH}). Then, it prepares the \emph{tmpSum}
vector so that the value \emph{tmpSum} is non-zero, only if there
are some bits enabled to the right of the \emph{lp} position (Listing:
\ref{listing:dcvEBarray:successor-part2}, Lines: \ref{code:successor:tmpSum-set-begin}
- \ref{code:successor:tmpSum-set-end}). In other words, the \emph{tmpSum}
is non zero only if there exists some successor candidate in some
of the current level's subtree. In such a case, the first loop is
interrupted (Listing: \ref{listing:dcvEBarray:successor-part2}, Line:
\ref{code:successor:first-inner-loop-break}) and the control goes
to the \emph{if} condition (Listing: \ref{listing:dcvEBarray:successor-part2},
Line: \ref{code:successor:last-level-check}), otherwise the level
variable \emph{cl} is decremented and the first inner loop starts
to examine the current node's parent (Listing: \ref{listing:dcvEBarray:successor-part2},
Line: \ref{code:successor:tmpSum-if-cond-start}). If the condition
\emph{cl = -1} is \emph{true}, then there are no successors in the
\emph{dcvEB} array and the procedure returns \emph{nil} (Listing:
\ref{listing:dcvEBarray:successor-part2}, Line: \ref{code:successor:no-further-successors-nil}).
The second inner loop (Listing: \ref{listing:dcvEBarray:successor-part2},
Lines: \ref{code:successor:second-inner-loop-begin} - \ref{code:successor:sec-inn-loop-end})
starts from checking the bit vector \emph{tmpSum}. The $0=$ \emph{tmpSum}
indicates that the successor candidate is removed after it has been
checked (Listing: \ref{listing:dcvEBarray:successor-part2}, Line:
\ref{code:successor:tmpSum-if-zero-restart-start}), hence the loop
is interrupted (Listing: \ref{listing:dcvEBarray:successor-part2},
Line: \ref{code:successor:tmpSum-if-zero-restart-stop}), and the
control goes back to the beginning of the outer loop \emph{while}
(Listing: \ref{listing:dcvEBarray:successor-part2}, Line: \ref{code:successor:first-outer-loop}).
If the successor is still there, the leftmost bit of \emph{tmpSum}
corresponding to the position of the first non-empty subtree is calculated
(Listing: \ref{listing:dcvEBarray:successor-part2}, Line: \ref{code:successor:most-left-bit}).
Then, the position in the bit vector \emph{bp} to the level position
\emph{lp} is transformed (Listing: \ref{listing:dcvEBarray:successor-part2},
Line: \ref{code:successor:lp-set}), and the variables \emph{pos},
\emph{ahol} and \emph{cAH} are appropriately updated (Listing: \ref{listing:dcvEBarray:successor-part2},
Lines: \ref{code:successor:pos-set} - \ref{code:successor:cAH-set}).
In particular , the value of \emph{cAH} is set to the subtree reference.
Hence, if the only element in the subtree was a successor candidate,
and unfortunately it was removed during the course of the second inner
loop, then the value of \emph{cAH} is \emph{nil} (Line: \ref{code:successor:cAH-eq-nil}).
In such a case, the algorithm must go up the tree and look for another
successor candidate i.e. go to the beginning of the outer loop \emph{while}
(Listing: \ref{listing:dcvEBarray:successor-part2}, Line: \ref{code:successor:goto-mark}).
Otherwise, the \emph{tmpSum} is set to \emph{cAH}'s bit vector (Line:
\ref{code:successor:tmpSum-update}) and the current level \emph{cl}
is incremented. The second inner loop ends when the current level
reaches the leaf level (Listing: \ref{listing:dcvEBarray:successor-part2},
Line: \ref{code:successor:sec-inn-loop-break}). At the leaf level,
the algorithm tries to fetch the key and data and, if it succeeds,
the successor element is returned. If not, the control goes back to
the beginning of the outer loop \emph{while}, and the algorithm begins
the journey up the tree. 

In the sequential case the \emph{successor()} method first traverses
the dcvEB array down (Listing: \ref{listing:dcvEBarray:successor-part1},
Line: \ref{code:successor:PreparePathToLeafCall}), next a bit up
(Listing: \ref{listing:dcvEBarray:successor-part2}, Lines: \ref{code:successor:first-inner-loop-begin}
- \ref{code:successor:first-inner-loop-end}), and then down again
(Listing: \ref{listing:dcvEBarray:successor-part2}, Lines: \ref{code:successor:second-inner-loop-begin}
- \ref{code:successor:main-while-end}). Thus, during their execution,
the \emph{successor()} method visits the nodes of the \emph{dcvEB
array} at most $3\dot{log_{n}\alpha}$ times. Thus, their sequential
running time is $O(3\cdot\log_{n}\alpha)=O(\log_{n}\alpha)$.

\section{Concurrency and dynamism\label{sec:Concurrency-and-dynamism}}

\subsection{Expanding and trimming\label{sub:Expanding-and-trimming}}

It is widely accepted that the trees in computer science are usually
drawn with the root at the top, and grow downwards instead of upwards.
\emph{Donald Knuth} in \cite{Knuth1973taoc} writes: \emph{\textquotedbl{}There
is an overwhelming tendency to make hand-drawn charts grow downwards
instead of upward (...) even the word 'subtree' (as opposed to 'supertree')
tends to connote a downward relationship\textquotedbl{}}. In this
context, the solution adopted in the \emph{dcvEB} array might be a
little unintuitive\footnote{Of course, in Mathematics, a \textquotedblleft tree\textquotedblright{}
is just an acyclic and connected graph. Thus, it can expand in all
directions. From this perspective, adding a new root over the old
one is not unusual.}. The tree of arrays, which in fact is the \emph{dcvEB} array, on
the one hand preserves the rule of the-root-at-the-top, on the other
hand, it grows in both directions: downwards and upwards. The \emph{dcvEB}
array's tree grows downwards if the key of the inserted element is
smaller than its current maximal capacity, and upwards if the key
of the inserted element exceeds the current capacity and the tree
needs to be expanded. Adding a new root above the current one makes
the old tree the leftmost subtree of the new root. Hence, the new
root summary, just after tree expansion, has the leftmost bit enabled.
Inserting the new element, which was the cause of expansion, enables
another bit in the root summary. The tree expansion increases the
size of the \emph{dcvEB} array (understood as the range of the allowable
keys) exponentially. Hence, adding $k$ levels above the current root
increases the size of a \emph{dcvEB} array $C_{new}=C_{old}\cdot n^{k}$
times, where $n$ is a summary size, $C_{old}$ - the old capacity
and $C_{new}$ - the new capacity. The \emph{dcvEB} array tree expansion
does not reorganize the old tree. Hence, all the existing elements
remain in place. Because the structure of the subtree does not change
the threads, which started to use the subtree before expansion, we
do not have to worry about the changed size of the tree. The variable
size of the tree does not affect the method of determining the position
of an element with the given key in a tree. Following the changed
size of a \emph{dcvEB} array (the height of the tree), the algorithm
also adapts to the new size of the array. Hence, $lp_{k}^{(h)}$ -
the level position computed for the tree of height $h$ and the fixed
key $i_{0}<n^{h}-1$ equals $lp_{k+r}^{(h+r)}$ - the level position
computed after the \emph{dcvEB} array tree expansion of $r$ levels,
where the level position $lp$ at the newly added levels is $0$ i.e.
$lp_{0}=\ldots=lp_{r-1}=0$. Since the expansion does not change the
structure of the existing tree, there is no reason to make any additional
blocking synchronization mechanisms due to the expansion itself. The
threads that start working before expansion also finish their work
within the old structure using the synchronization scheme as is presented
in \cite{Kulakowski2014acve}. The new threads that start working
after the expansion use the new structure, however, within the subtree
resulting from the previous tree, they synchronize with the ``old''
threads using the locks available within the subtree. Hence, the main
role of synchronization in the case of the tree expansion is to provide
the new threads with the latest consistent information about the \emph{dcvEB}
array tree. For this reason, the vital \emph{dcvEB} array information,
such as \emph{root} reference, \emph{height}, and the associated \emph{size,}
are kept in the atomic global variable \emph{ap} (\emph{ArrayParam}
structure). That is why every interface method starts from fetching
the current array parameters (Listings: \ref{listing:cvEBarray:Insert},
\ref{listing:dcvEBarray:Get}, \ref{listing:dcvEBarray:Delete}, \ref{listing:dcvEBarray:successor-part1}
Lines: \ref{code:Insert:ap2cap}, \ref{code:get:ap2cap}, \ref{code:delete:ap2cap},
\ref{code:successor:ap-fetch}). The \emph{dcvEB} array expansion
is implemented as part of the \emph{insert() }implementation (Listing:
\ref{listing:cvEBarray:Insert}). The auxiliary method \emph{grow()}
first prepares the new top of the \emph{dcvEB} array tree, then the
\emph{insert()} method tries to atomically replace the current \emph{ArrayParam}
set with the new one prepared by \emph{grow()}. The \emph{insert()}
method managed to set the new top of the tree. The gray triangle represents
the tree before expansion. 

Most methods do not interfere with \emph{insert()} as regards extending
the tree. This group includes queries such as \emph{get()}, \emph{successor()},
\emph{predecessor()}, \emph{min()} and \emph{max()}. If the expansion
occurs after they fetch their own \emph{AP} copy, the result they
return will just not take into account the new elements, for which
the key is greater than the size of the structure before expansion.
Hence, the synchronization with respect to these methods can be limited
to the nonblocking operations on atomic variables. Unfortunately,
in the case of \emph{delete()} a \emph{dcvEB} array tree expansion
would be easily disrupted by the trimming procedure.

It is possible that, if there were no additional synchronization mechanism
a global read-write lock \emph{apLock},\emph{ }(Listings: \ref{listing:cvEBarray:Insert},
\ref{listing:cvEBarray:topTrim}). In the absence of this mechanism
it is possible that between fetching the new array parameters (Listing:
\ref{listing:cvEBarray:Insert}, Line: \ref{code:Insert:ap2cap})
and read-locking the root (Listing: \ref{listing:cvEBarray:Insert},
Line: \ref{code:Insert:root:rLock}) the top of the array would be
trimmed. In such a case, \emph{insert()} would use the root \emph{AH}
that was in fact removed from the tree, thus an insertion would be
ineffective. Hence, the aim of the lock combinations (\emph{apLock}
read lock and \emph{AH} read lock) used in both \emph{insert()} and
\emph{topTrim()} is to prevent removal of the node when it is examined
due to the insertion procedure. Both locks support reader-writer semantics.
Hence, many different inserts but only one delete can be handled at
the same time. 

Calling \emph{delete()} during the expansion of the tree might also
cause another problem that may lead (if not handled) in the long time
perspective to performance deterioration. Namely, \emph{delete()}
first goes towards the leaf, then removes it, and next it tries to
propagate the delete information to the higher levels of the tree.
If the tree is extended after \emph{delete()} fetches the current
\emph{AP} snapshot (Listing: \ref{listing:dcvEBarray:Delete}, Line:
\ref{code:delete:ap2cap}), then the highest reachable node for delete
is the root of the tree before expansion. Hence, \emph{delIntern()}
is not able to propagate delete information up to the new root and
ends earlier within a subtree. In such a case, \emph{delete()} may
leave the path starting from the new root (and ending in the old root),
which is composed of nodes containing false information indicating
that their subtrees are non-empty. This issue can be solved in a few
different ways. One of them could be to relax the delete operation
and use query methods like \emph{get()} or \emph{successor()} to clean
up the tree. Another, proposed in this article, is to repeat \emph{deleteClean()}
(Listing: \ref{listing:dcvEBarray:Delete}, Line: \ref{code:Delete:deleteClean})
- the procedure, which discovers an undeleted path's residues and
removes them when needed. The method \emph{deleteClean()} is very
similar in implementation to \emph{delete()} itself. However, other
than \emph{delete(),} it does not remove the element, but instead
it tries to confirm that there is no element with the given key in
the tree. If there is no such element, but there are some \emph{AHs}
in the tree leading to them and only to them, then such nodes are
removed and appropriate bits are disabled. The implementation of the
functionality of \emph{deleteClean()} boils down to small changes
in the code of the \emph{makePath() }and \emph{delIntern()} methods.
Hence, taking into account that the \emph{Java} code of the presented
solution is publicly available, the pseudocode of \emph{deleteClean()}
is not thoroughly analyzed in the paper. An optimal number of \emph{deleteClean()}
calls is discussed later, when the progress condition is considered. 

As the \emph{dcvEB} array grows upwards, it is trimmed from the top.
The trimming routine is located at the end of the \emph{delete()}
method. It checks whether the current root has only one child at position
$0$, and if so, it attempts to remove the root. Hence, the root can
be trimmed if the maximal element stored within the structure is smaller
than $size/n$, where $size$ means the current size of the structure
(the largest key that can be inserted into the dcvEB array without
triggering its extension), and $n$ is the length of the summary bit
vector. Trimming is implemented as the inverse of expansion. The only
difference is that the \emph{insert()} method tries to insert all
the required nodes at once, whilst \emph{delete()} trims elements
one by one. The root element after trimming is removed from the tree,
and its only child is promoted to the new root element. The trimming
mechanism is completely neutral for \emph{get()}. This is partly a
merit of \emph{Java}, where, of course, nodes can be detached from
the tree, but they are not destroyed, just as in \emph{C++} or \emph{C}.
Hence, as only \emph{get()} obtains its own reference to the \emph{AP}
structure (Listing: \ref{listing:dcvEBarray:Get}, Line: \ref{code:get:ap2cap}),
and thus, the reference to the root, then regardless of whether any
visited node is detached from the tree, \emph{get()} may always continue
traversing the structure. Similarly, the other query methods are not
affected by the trimming. For instance, \emph{successor()}, when it
obtains the root snapshot, creates the path snapshot from the root
to the lowest existing node on the path leading to the element with
the given key (Listing: \ref{listing:dcvEBarray:successor-part1},
Line: \ref{code:successor:PreparePathToLeafCall}). The path is kept
in the two auxiliary arrays \emph{ahol} and \emph{pos}. Thus, if some
of the vertices stored in \emph{ahol} are removed from the tree, they
are not removed from \emph{ahol}, thus the \emph{successor()} algorithm
still has access to them. Hence, if there is a need to traverse the
tree upwards, \emph{successor()} is always able to do that. On the
other hand, if some of the visited nodes have been actually removed
from the tree, this also implies that there are no further successor
candidates available, and \emph{successor()} may finish its task earlier.
For the same reason, trimming does not affect \emph{delete()} either.
Since, during the passage down, \emph{delete()} creates its own \emph{ahol}
array, it has no problems with propagation of the deletion status
up. Of course, if some of the nodes referenced in \emph{ahol} are
removed (trimmed) from the tree, \emph{delete()} might stop status
propagation immediately after encountering such a node. This observation,
as well as an analogous situation in the case of the successor search,
can also be a subject for further optimization. 

Two or more different \emph{topTrim()} calls are synchronized with
each other by using \emph{apLock} writer lock. On the other hand,
for the purpose of mutual \emph{topTrim()} synchronization, a nonblocking
mechanism seems to be sufficient. Hence, the \emph{apLock} usage in
this place is caused only by the necessity of protecting \emph{insert()
}against trimming. The writer locks used here (Listing: \ref{listing:cvEBarray:topTrim}
Line: \ref{code:toptrim:apLock}) provide \emph{topTrim()} exclusive
access to the top of the structure, which could be potentially dangerous
for the concurrent performance of the structure. However, in practice,
the need to trim the top of the \emph{dcvEB} array tree is not frequent,
thus with the performance issue in mind, it is better first to check
whether the trimming is needed (Listing: \ref{listing:cvEBarray:topTrim},
Lines: \ref{code:toptrim:while:begin} - \ref{code:toptrim:trim-stop}),
and if so, lock the top of the structure, and before trimming, check
once again (Listing: \ref{listing:cvEBarray:topTrim}, Line: \ref{code:toptrim:if:trimming-needed})
whether the trimming is actually needed.

\subsection{Dynamic adding and removing elements\label{sub:Dynamic-adding-and}}

In addition to extending and trimming the root, the \emph{dcvEB array}
tree also adds and removes elements below the root. This is similar
to the behaviour known from the classical trees, where inserting or
removing data causes the creation or deletion of appropriate nodes.
In the presented solution, it is assumed (due to the desire to avoid
separate handling of the uninitialized root case) that there is always
at least one node within the tree. This means that the root node exists,
even if there are no other elements in the tree. The \emph{insert()}
method creates the new element when it turns out that there is no
\emph{AH} at the specified position (Listing: \ref{listing:cvEBarray:Insert},
Line: \ref{code:Insert:subtree-AH-creation}). Adding a child \emph{AH}
into the parent array may interfere with another insert activity.
Hence, to avoid overwriting one \emph{AH} by another \emph{AH}, the
parent's array update is implemented as a \emph{CAS} instruction (Listing:
\ref{listing:cvEBarray:Insert}, \ref{code:Insert:assign-to-a-parent-array}).
Then, if the first thread wins and successfully updates the parent
\emph{AH's} array, the second thread just retrieves the winning \emph{AH}
(Listing: \ref{listing:cvEBarray:Insert}, Line: \ref{code:Insert:cAHfinalValue})
and continues the insertion procedure with them. It is worth noting
that the whole operation is protected in the same way as the summary
bit-vector update, i.e. by the \emph{AH's} reader lock (Listing: \ref{listing:cvEBarray:Insert},
Line: \ref{code:Insert:rLock}). Thus, there is no risk that the newly
updated \emph{AH} would be removed by \emph{delete()} (after execution
Line: \ref{code:Insert:assign-to-a-parent-array}), and \emph{insert()}
could fetch a \emph{nil} value (Listing: \ref{listing:cvEBarray:Insert},
Line: \ref{code:Insert:cAHfinalValue}). A more detailed analysis
of the synchronization scheme used here can be found in \cite{Kulakowski2014acve}. 

Since operations related to the memory allocation and deallocation
involve operating system function call, they are usually time-consuming.
Therefore, for the purpose of this algorithm a kind of lazy approach
has been adopted. Therefore, the \emph{AHs} are not removed from the
tree immediately after they become empty (leaves are considered empty
when the fields' data and index are set to \emph{nil}, whilst other
nodes are empty if their summary bit-vector equals $0$). Instead,
if some non-leaf node becomes empty, it actually removes all its children
\emph{AH}s from the memory. Therefore, the operation of deleting elements
can be considered as composed of two phases: the first one - logical
removing - when data and index are set to \emph{nil} or appropriate
position in the parent's summary bit-vector is zeroed, and the second
one - physical removing - when the \emph{AH's} reference is physically
removed from the parent's array and the \emph{AH} record is actually
removed from the memory. Since the \emph{AH} record is physically
removed (Listing: \ref{listing:dcvEBarray:internalDelete}, Lines:
\ref{code:Delete:physically-remove} - \ref{parentAH.summary.summary-check-end})
only when it has no siblings, the physical removing is likely to occur
less frequently than the logical removing. Both logical and physical
removing use the same synchronization scheme as presented in \cite{Kulakowski2014acve}.
Thus, the modified elements are always exclusively held by the thread
performing deletion.

\subsection{Successor search strategy\label{sub:Successor-search-strategy}}

The working scheme of the \emph{successor()} method is composed of
three phases. During the first one (Listing: \ref{listing:cvEBarray:PreparePathToLeaf}),
the algorithm goes as far as possible towards the element indexed
by the given key. If the element exists, it returns them. If not,
the control goes to the second phase (Listing: \ref{listing:dcvEBarray:successor-part2},
Lines: \ref{code:successor:first-inner-loop-begin} - \ref{code:successor:first-inner-loop-end})
in which the algorithm retracts until the next nonempty subtree is
found. Then, during the third phase (Listing: \ref{listing:dcvEBarray:successor-part2},
Lines: \ref{code:successor:second-inner-loop-begin} - \ref{code:successor:sec-inn-loop-end})
the algorithm goes down towards the element which is minimal within
the detected nonempty subtree. If the third phase fails (the desired
element can be removed in the meantime), then the control goes to
the second phase and the algorithm starts to go upwards. The second
and third phases are repeated as many times as needed. The main difference
in comparison with \cite{Kulakowski2014acve} is that, in the case
of the necessity to repeat the second and the third phase, the second
phase starts exactly from the same point where the third phase has
been stopped. Thus, with each search failure the \emph{successor()}
method tries to look further for the next possible successor candidate.
Therefore, the number of repetitions of phase two and three is naturally
limited by the size of the \emph{dcvEB} array. 

Thanks to the adopted strategy, the successor element will always
be found if it remains in a \emph{dcvEB} array long enough. Thus,
let \emph{key} be the index of element \emph{x} whose successor we
are looking for, and let \emph{succ }be the index of \emph{y} - some
successor of \emph{x, }such that \emph{key} < \emph{succ}. In such
a case, if during processing the second part of the \emph{successor()}
method (Listing: \ref{listing:dcvEBarray:successor-part2}) \emph{y
}is not removed from the \emph{dcvEB} array, then the maximal possible
key of the next successor of \emph{x} is \emph{succ}. In other words,
when the search algorithm detects the subtree (phase 2) containing
\emph{y} as the minimal element, then the algorithm seamlessly (i.e.
without failures) reaches \emph{y}. The current algorithm always returns
the successor if it is available during the whole course of the \emph{successor()}
method. In the previous implementation \cite{Kulakowski2014acve}
there was a small chance that the successor would not be found and
\emph{successor()} would return \emph{nil}. On the other hand, the
previous implementation limits the number of failures that can be
safely handled by the \emph{successor()}, whilst currently, the allowable
number of failures is limited only by the current size of the structure.
This raises the question of the actual concurrent running time of
the current \emph{successor()} implementation. As will be shown in
the next section (Sec. \ref{sub:Optimal-concurrent-successor}), the
concurrent running time estimation is worse than in the sequential
case. Fortunately, the tests carried out indicate rather high overall
efficiency of the structure rather than its susceptibility to the
interferences and thereby performance deterioration.

\subsection{Successor search concurrent running time\label{sub:Optimal-concurrent-successor}}

The sequential running time of all the methods presented in the article
is the same as in \cite{Kulakowski2014acve}, and equals $O(log_{n}\alpha)$,
where $n$ is the size of the summary bit vector, whilst $\alpha$
is the current size of the \emph{dcvEB} array. The detailed arguments
presented previously, with only minor amendments, also fit the \emph{dcvEB}
array. The concurrent running time estimation is much more complex,
because, besides the code structure, different kinds of concurrent
interactions, such as blocking synchronization, need to be taken into
account. Fortunately, some of the \emph{dcvEB} array methods discussed
in this article use only non-blocking synchronization mechanisms.
These methods are: \emph{get()} and \emph{successor()}. Since the
\emph{get()} method only checks the presence of one, well-defined
element in the array, if the check fails (no matter when the element
has been removed), it returns \emph{nil}. Hence, its concurrent running
time estimation is not affected by the interferences with other concurrently
executed methods. Thus, the concurrent running time of \emph{get()}
does not change and is $O(log_{n}\alpha)$. 

In contrast, in the case of \emph{successor()}, despite the non-blocking
synchronization, every deletion of a successor candidate may increase
the overall concurrent running time of the method. In the worst case
scenario, the value of the \emph{successor()} method is called for
input argument $0$, and the next greater element has the key $1$.
Then, if the element indexed by $1$ is deleted just after \emph{successor()}
reaches the end of the loop (Listing: \ref{listing:dcvEBarray:successor-part2},
Line: \ref{code:successor:first-inner-loop-end}), then the deletion
has been discovered (Listing: \ref{listing:dcvEBarray:successor-part2},
Line: \ref{code:successor:deletion-check}) and the next iteration
will be initiated (Listing: \ref{listing:dcvEBarray:successor-part2},
Line: \ref{code:successor:next-iteration-goto}). This may lead to
alteration of the subsequent removal and search attempts. The worst
case scenario described above may occur $\alpha-1$ times, forcing
the \emph{successor()} method to check every single position in the
\emph{dcvEB} array. Hence, in the worst case, the concurrent running
time of \emph{successor()} is $O(\alpha)$. A natural question arises
whether this slightly disappointing result can be improved. The easiest
solution (but without any guarantee that the successor will be found
even if it exists in the tree) is to limit the possible number of
iterations of phases 2 and 3 (the loop \emph{while} Listing: \ref{listing:dcvEBarray:successor-part2},
Lines: \ref{code:successor:first-outer-loop} - \ref{code:successor:main-while-end})
by some empirically chosen constant. In such a case, the user must
accept that (probably) very rarely the \emph{successor()} method would
fail and never return the correct value\footnote{In fact, this approach provides a ``limited warranty'' to find a
successor, i.e. if the iteration number is limited by e.g. $1000$
and the \emph{successor()} method is called with $\alpha$ on its
input, then (even in the worst case scenario) there is a guarantee
that if the successor exists during the \emph{successor()} call, then
it will be found if only $\beta-\alpha\leq1000$, where $\beta$ is
the key of successor. }. Despite the moderately good theoretical concurrent running time
estimation, the current solution performs very well in practice. Conducted
tests for the random data (Section \ref{sec:Experimental-results})
seem to suggest that the \emph{successor()} average running time is
closer to $O(log_{n}\alpha)$ than $O(\alpha)$. 

There are also some theoretical arguments that may indicate in favour
of $O(log_{n}\alpha)$. For simplicity, let us assume that the \emph{dcvEB
array} is full (without \emph{nil} values) and every deleted element
is re-inserted into the table in a short time after removal. Hence,
without making a big mistake, it can be assumed that the successor
of $s_{k}$ is $s_{k+1}$ for $k=1,\ldots,m$. Thus, the collision
may happen, if at roughly the same time, \emph{successor() }processes
$s_{k}$ and \emph{delete() }removes\emph{ $s_{k+1}$. }Assuming that
both: \emph{successor()} and \emph{delete()} process $m$ randomly
selected indices: $(s_{1},\ldots,s_{m})$ and $(d_{1},\ldots,d_{m})$
at the same and equal time intervals, the likelihood\footnote{It is enough to assume that at the certain moment of time $d_{r}$
is fixed, whilst the index $s_{k}$ is selected as one out of $1,\ldots,N$. } of the simultaneous execution of \emph{delete($d_{r}$) }and\emph{
successor($s_{k}$ ) }where $d_{r}=s_{k}+1$ is $1/\alpha$. Hence,
the execution of $m$ consecutive \emph{successor()} and \emph{delete()}
calls may result in $m/\alpha$ collisions. Since every collision
entails additional tree traversal by \emph{successor(),} the total
expected concurrent running time of $m$ \emph{successor()} calls
is $T(\alpha,m)=mO(\log_{n}\alpha)+(m/\alpha)O(\log_{n}\alpha)$.
Thus, the amortized concurrent running time of a single \emph{successor()}
call is $T(\alpha,m)/m=(1+1/\alpha)O(\log_{n}\alpha)$. In most cases
$\alpha$ is large, for that it is safe to assume that $T(\alpha,m)/m\approx O(\log_{n}\alpha)$.
Similarly, the amortized concurrent running time of a single successor
call for $h$\emph{ }delete threads running in parallel is $T(\alpha,m)/m=(1+h/\alpha)O(\log_{n}\alpha)$.
Thus, as long as $h$ is significantly smaller than $\alpha$ it still
holds that $T(\alpha,m)/m\approx O(\log_{n}\alpha)$. At the expense
of increasing the complexity, the above reasoning can be adapted to
the \emph{successor()} algorithm as presented in the article.

\subsection{Correctness}

As with other concurrent data structures, the \emph{dcvEB} array should
be considered while bearing in mind the concurrent objects specificity\cite{Lea1999cpij,Herlihy2008taom}.
In particular, correctness is discussed in terms of data consistency
(quiescent and sequential), linearizability \cite{Wing1990lacc},
and the progress condition. Although these problems were already discussed
in \cite{Kulakowski2014acve}, some issues need to be revisited due
to the changes made to the original idea. One of them is linearizability.
The proposed enhancements are of two types: the first group concerns
growing and shrinking the \emph{dcvEB} array tree, whilst the second
concerns allocation and deallocation memory for \emph{AHs}. Bearing
in mind the principle \cite{Herlihy2008taom} according to which every
linearizable method call should appear to take effect instantaneously
at some moment between its invocation and response, it is clear that
the first group of enhancements do not affect linearizability property.
In the case of \emph{insert()} and \emph{delete(),} natural linearization
points are unlock calls just after setting the values into the leaf
\emph{AH} (Listing: \ref{listing:cvEBarray:Insert}, Line: \ref{code:Insert:previousAH:unlock}),
and (Listing: \ref{listing:dcvEBarray:internalDelete}, Lines: \ref{code:Delete:while2:linearization_point_1}
- \ref{code:Delete:while2:linearization_point_2}). Since the code
responsible for the array growth (inserting) or trimming (deleting)
is not directly involved in setting or unsetting the new values in
\emph{AHs,} then the linearizability property is not affected by the
first group of enhancements. The second group of enhancements do not
change the logic of inserting and deleting elements either. This is
because they relate to the memory management rather than deleting
the elements (elements cease to be available to the user at the same
time as previously, i.e. the element is considered to be deleted when
its value in the leaf \emph{AH} is overwritten by \emph{nil} (Listing:
\ref{listing:dcvEBarray:internalDelete}, Line: \ref{code:internalDelete:dataOverwritten})).
In other words, despite the fact the memory is not immediately freed
after the \emph{delete()} call (due to the lazy approach to the memory
deallocation), the \emph{delete()} method logically takes effect immediately.

Similarly, all the query methods, including \emph{get()} and \emph{successor(),}
are also linearizable i.e. they take effect before the query value
is returned. Therefore, the \emph{dcvEB} array is linearizable, hence,
it is also quiescently and sequentially consistent \cite{Herlihy2008taom}.

Another important criterion of concurrent object correctness is the
progress condition. As has been shown in \cite{Kulakowski2014acve},
to preserve the progress condition the fair locking strategy needs
to be used. Since the general synchronization scheme remains unchanged,
a potential risk to the progress condition may come from the introduced
enhancements. Especially dangerous are those loops in which fulfillment
of the stop condition is not obvious. The first such loop is responsible
for the \emph{dcvEB} array growing (Listing: \ref{code:Insert:size-check},
Lines: \ref{code:Insert:while--growing-start} - \ref{code:Insert:while-stop}).
The loop executes until the size of the array \emph{cAP.size} is greater
than the requested \emph{key}. To prove that the iteration of the
loop will end, note first that in every turn of the loop \emph{cAP.size}
can only grow. That is because \emph{cAP.root} is reader-locked (Listing:
\ref{listing:cvEBarray:Insert}, Line: \ref{code:Insert:root:rLock}),
hence the method \emph{delete()} is not able to trim the \emph{dcvEB}
\emph{array} in the meantime. Secondly, let us note that in every
turn of the loop \emph{cAP.height} increases by at least one. Although
within the current loop \emph{insert()} tries to increase the array
as far as necessary, it is possible that it loses the race with another
thread, which increases the array height only by one. 

Another important loop with an unobvious end-condition is connected
with the cleaning after deletion (Listing: \ref{listing:dcvEBarray:Delete}.
Lines: \ref{code:Delete:while-with-deleteClean-begin} - \ref{code:Delete:while-with-deleteClean-end}).
Although the number of possible iterations is limited by the \emph{maxRep}
constant, the question arises as to how many iterations are sufficient,
i.e. how large the \emph{maxRep} constant should be. To answer this
question, let us recall that the loop is introduced to prevent incomplete
deletion, which may happen if, during the element removal, the tree
is expanded up by the \emph{insert()} method. In such a case, since
\emph{delete()} knows only the root before expansion, then it is able
to propagate the delete information only up to this old root. Hence,
if some nodes placed above the old root also need to be removed, the
solution proposes a re-run of the \emph{deleteClean}() method. It
starts from the new root and checks whether for all the nodes on the
path from the (new) root and the requested element, information contained
in \emph{AH's} fields summary and \emph{AH's} array are mutually consistent.
Hence, the question about the optimal value of \emph{maxRep} (Listing:
\ref{listing:dcvEBarray:Delete}, Line: \ref{code:Delete:while-with-deleteClean-begin})
is the question about how many times in a row there may be a situation
that the \emph{dcvEB} array tree will be expanding up when the \emph{delIntern()}
or \emph{deleteClean()} operations are ongoing. The answer depends
on the domain of the array's key. For instance, if the array key is
\emph{Java's Integer}\footnote{The maximal integer and the long value in \emph{Java are} $MAX\mbox{\_}INT=2^{31}-1$,
$MAX\mbox{\_}LONG=2^{63}-1$.}, then assuming the length of the bit-vector as $64$, the height
of the \emph{dcvEB} array tree that could hold an object with the
key equal to the maximal representable integer is $6=\left\lceil \log_{64}(2^{31}-1)\right\rceil $.
For \emph{Java's Long,} the sufficient height is $11=\left\lceil \log_{64}(2^{63}-1)\right\rceil $.
Hence, for an array indexed by \emph{Java's Integer,} it is enough
to set \emph{maxRep} to $6$, or to $11$ if an array is indexed by
\emph{Java's Long}. Of course, the exact value of \emph{maxRep} depends
on the specific data types and hardware platform and needs to be re-calculated
for every specific implementation.

There is also one ``hidden'' loop, mentioned in \cite{Kulakowski2014acve},
which is not explicitly shown in pseudo-code, although it is important
for practical implementation. This loop is connected with atomic update
of the summary bit-vector within the \emph{insert()} method (Listing:
\ref{listing:cvEBarray:Insert}, Line: \ref{code:Insert:setBit}).
Since, in most of the programming languages including \emph{Java,}
bit operations are not atomic, this update has to be implemented according
to the scheme: atomic read, modify, atomic compare and set (Listing:
\ref{listing:cvEBarray:atomic-bit-vector-update}). 

\setcounter{algocf}{11}
\SetAlFnt{\small\sf}
\IncMargin{1em}
\begin{algorithm}

\texttt{while(true) }

\texttt{~~s $\leftarrow$ cAH.summary;\label{code:bitvector-update:fetch-summary}}

\texttt{~~if $0_{n}0_{n-1}\ldots0_{n-lp+1}1_{n-lp}0_{n-lp-1}\ldots0_{1}$
\label{code:bitvector-update:first-if-close-start}}

\texttt{~~~~$AND_{bit}$~s > $0$ \label{code:bitvector-update:first-if-close-end}}

\texttt{~~~~break;\label{code:bitvector-update:first-if-cond}}

\texttt{~~if CAS(cAH.summary, s, s $OR_{bit}$}

\texttt{~~~~$0_{n}0_{n-1}\ldots0_{n-lp+1}1_{n-lp}0_{n-lp-1}\ldots0_{1}$)\label{code:bitvector-update:cas}}

\texttt{~~~~break;}

\texttt{end}

\NoCaptionOfAlgo
\caption{\textbf{Listing \thealgocf:} Atomic bit-vector update scheme} \label{listing:cvEBarray:atomic-bit-vector-update}

\end{algorithm}
\DecMargin{1em}

In order to prove that the bit-vector update scheme presented above
meets the progress condition, it is necessary to note that \emph{cAH}
is reader-locked within the \emph{insert()} method. Thus, the only
concurrent modifications coming from other threads that may occur
within the code (Listing: \ref{listing:cvEBarray:atomic-bit-vector-update},
Lines: \ref{code:bitvector-update:fetch-summary} - \ref{code:bitvector-update:cas})
may enable another bit in \emph{cAH.summary} (but not disable them).
Since \emph{cAH.summary }is modified (Listing: \ref{listing:cvEBarray:atomic-bit-vector-update},
Line: \ref{code:bitvector-update:cas}) only if the requested bit
is not enabled (Listing: \ref{listing:cvEBarray:atomic-bit-vector-update},
Line: \ref{code:bitvector-update:first-if-cond}) then the \emph{CAS}
instruction may fail in the worst case $n$ times in a row, where
$n$ is the size of the bit vector. So, at most after $n$ iterations
\emph{cAH.summary} contains only enabled bits, thus the result of
the \emph{if} condition (Listing: \ref{listing:cvEBarray:atomic-bit-vector-update},
Lines: \ref{code:bitvector-update:first-if-close-start} - \ref{code:bitvector-update:first-if-close-end})
is always true. As a result, at most after $n$ iterations the loop
is stopped. Hence, the progress condition is preserved.

\section{Experimental results\label{sec:Experimental-results}}

The experimental implementation of the test application together with
the \emph{dcvEB array} was written in \emph{Java} 7 and has been tested
on an isolated test station Intel{\footnotesize\raisebox{1pt}{\textregistered}}~Core\texttrademark~i7-3930K
(6 cores, 12 threads, 3.8 GHz) processor with 8 GB of operating memory.
As in the case of the \emph{cvEB} \emph{array,} the results achieved
by its successor, the \emph{dcvEB array,} are very promising. In many
cases, the new structure turns out to be faster than the compared
alternatives. Of course, the presented results are indicative and
do not pretend to be a ranking or review. 

Since one of the advantages of the \emph{dcvEB array} is to support
all the dynamic set's methods mentioned in \cite[p. 230]{Cormen2009ita},
one of the major challenges was to find an appropriate concurrent
dynamic set implementation, which in addition to the standard \emph{get()},
\emph{delete()} and \emph{insert()} also provides \emph{successor()}
and \emph{predecessor(). }An experimental implementation of the \emph{dcvEB
array} has been written in \emph{Java. }Since sometimes even small
implementation details may affect the overall application performance,
it was equally important to find such dynamic set solutions written
in \emph{Java} that have been identified by the authors as reference
solutions. 

Besides the \emph{dcvEB array,} \emph{SnapTree} \emph{Map} \cite{Bronson2010bsSIMPL},
\emph{ConcurrentSkipListMap} \emph{\cite{Herlihy2008taom}, }non-blocking
k-ary search tree \cite{Brown2011nbksSIMPL} and the synchronized
\emph{java.util.TreeMap} were selected for the tests. The structure
proposed by \emph{Bronson} \emph{et al}. was selected for testing
because of the publicly available \emph{Java} implementation provided
by the authors\footnote{\href{https://github.com/nbronson/snaptree}{https://github.com/nbronson/snaptree}},
and the presence of support for \emph{successor()} and \emph{predecessor()}
(via Java's \emph{ConcurrentNavigableMap} interface). The second structure,
\emph{ConcurrentSkipListMap},\emph{ }as it is implemented in recent
\emph{Java} distributions was a natural candidate for comparison.
In \cite{Herlihy2007asos} \emph{Herlihy} et al. wrote about \emph{ConcurrentSkipListMap
}that ``\emph{(...) written by Doug Lea based on work by Fraser and
Harris \cite{Fraser2004plf} and released as part of the JavaTM SE
6 platform, is the most effective concurrent SkipList implementation
that we are aware of''. }The structure also implements \emph{successor()}
and \emph{predecessor()}. Moreover, its inclusion into the standard
\emph{Java 6} platform indicates its optimality. 

The non-blocking k-ary search tree (\emph{LockFree9ST} class\footnote{\href{http://www.cs.utoronto.ca/~tabrown/ksts/}{www.cs.utoronto.ca/$\sim$tabrown/ksts/}})
\cite{Brown2011nbksSIMPL} is a new efficient tree structure modeled
on \cite{Ellen2010nbbsSIMPL} with the publicly available \emph{Java}
implementation. It does not support the \emph{successor()} and \emph{predecessor()}
operations. Thus, for the test purposes, the successor operation has
been implemented as the iterative \emph{get()} method calls for subsequent
indices. The iteration ends when \emph{get()} returns a non-null element
or the iteration reaches the largest possible index that can be stored
in the tree. In general, such an approach is not effective, especially
when the indices are distributed sparsely. Fortunately, during the
tests the indices are distributed fairly densely, thus, one may expect
that the performance deterioration resulting from using this ``naive''\emph{
}successor search strategy is not too high. 

In contrast to the \emph{ConcurrentSkipListMap} structure, \emph{TreeMap
}is not even a concurrent object implementation. It is a globally
synchronized sequential structure, which is included into the tests
in order to show the difference between concurrent and sequential
approaches.

In testing, there are four groups of threads: \emph{getters()}, \emph{inserters()},
\emph{removers()}, and \emph{successors()}\footnote{Due to the symmetrical similarity with \emph{successor(), }the \emph{predecessor()}
method has not been subject to testing.}. Each getter thread repeatedly calls the method \emph{get()}, inserter
thread - \emph{insert()} etc. Every thread has to perform the same
$z$ number of calls. The input parameter $key$ is randomly chosen
from $K\subseteq[0,\ldots,x-1]$, where $x$ is a test run parameter.
The number of threads in groups is denoted by the letters g, i, r
and s. For instance, g $=3$, s $=5$ means that in the given test
run three getters and five successor searchers are involved. Since
the number of operations per thread is fixed, the method performance
is measured as an execution time according to the principle, the lower
the value, the better the result. The execution time is measured per
thread, thus the time of the whole test run is the arithmetic mean
of the execution time of each thread involved. In other words, the
thread execution time is the time needed by the thread to make $z$
calls of the given method. 

\begin{figure}[h]
\begin{centering}
\includegraphics[scale=0.26]{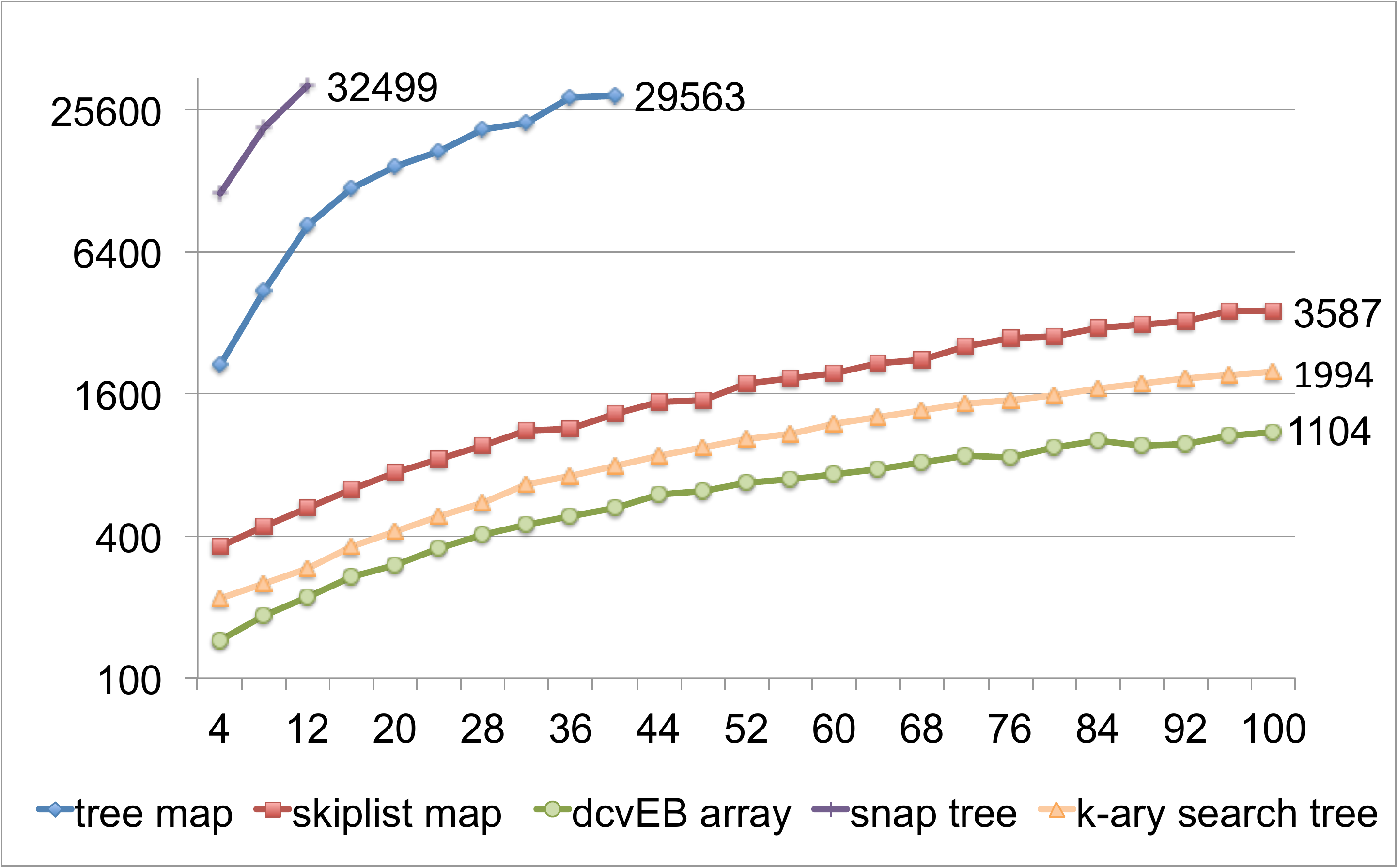}
\par\end{centering}

\caption{The \emph{dcvEB} array overall performance test results ($g=i=r=s=1,2\ldots,25$,
$z=500000$ and $m=500000$)}
\label{fig:test1}
\end{figure}

As before, the first test (Fig. \ref{fig:test1}) measures how the
overall thread execution time grows, when the number of threads in
each group is the same and increases. In the first test run there
are four threads $g=i=r=s=1$, in the second one eight threads $g=i=r=s=2$
and, finally, in the last one $100$ threads $g=i=r=s=25$. Every
thread performs $z=500000$ method calls. The results are averaged,
so that the bad test result of one method can be compensated by the
good result of another method.

The best result has been achieved by the \emph{dcvEB array}. Its averaged
thread execution time gets $1104$ ms. and is approximately $1.8$
times better than the second result belonging to \emph{LockFree9ST}.
Both \emph{TreeMap} and \emph{SnapTree} are far behind the two previous
structures. Such a structure of results demonstrates that the improvements
made to the original \emph{cvEB} \emph{array} do not affect its overall
performance, and, on the contrary, they seem to be even better than
in \cite{Kulakowski2014acve}. 

In addition to the general performance comparison of the structures,
it is interesting to compare the performance of the particular methods.
As in \cite{Kulakowski2014acve}, the results were averaged, so that
the particular number in milliseconds denotes the average time required
by a getter, an inserter, a deleter and a successor searcher thread
to execute half a million calls of the given method. 

\begin{figure}[h]
\begin{centering}
\includegraphics[scale=0.24]{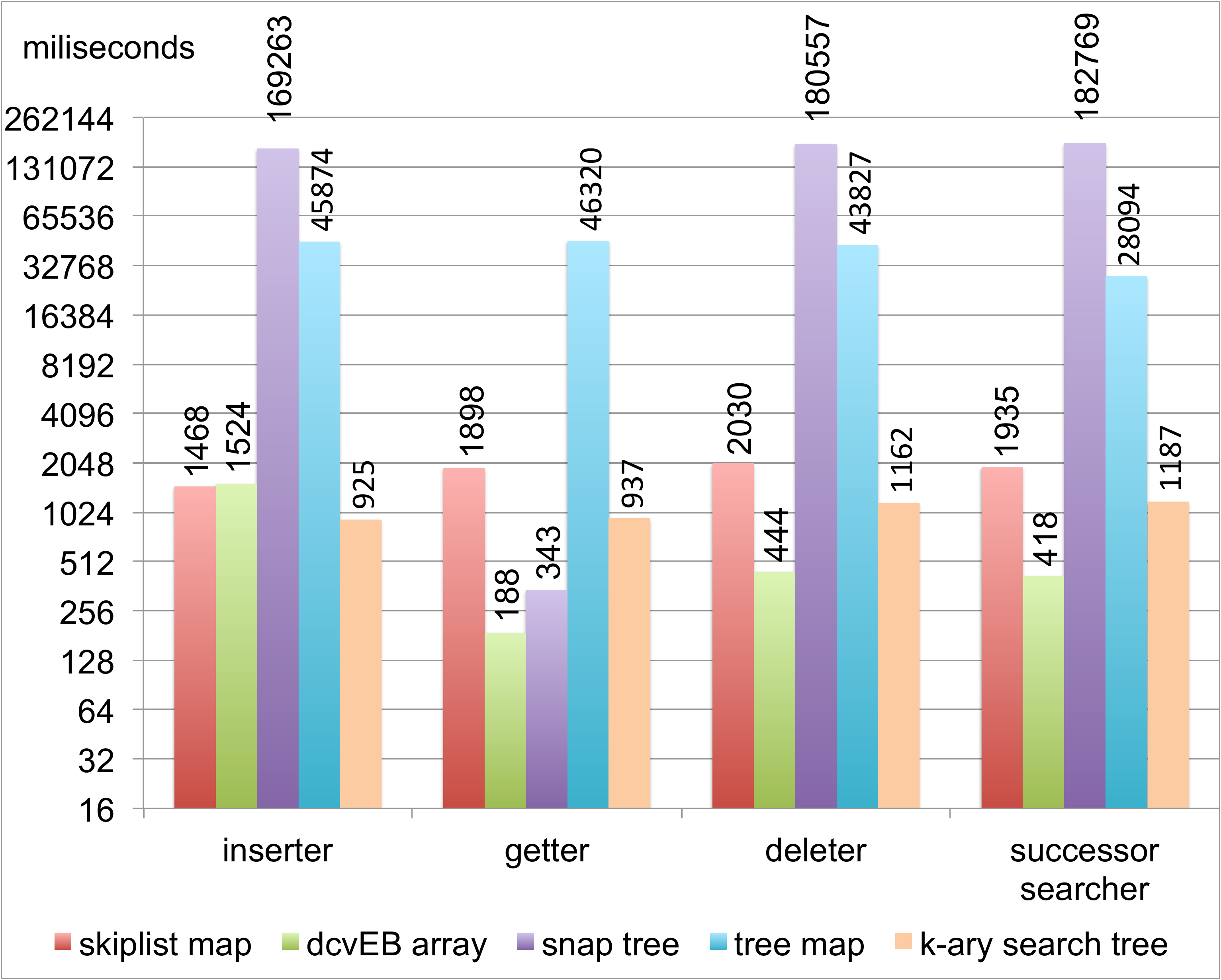}
\par\end{centering}

\caption{The \emph{dcvEB} array method performance test results ($g=i=r=s=1,2\ldots,25$,
$z=500000$ and $m=500000$).}
\label{fig:test1:methods}
\end{figure}

The obtained results (Fig. \ref{fig:test1:methods}) show that the
performance of each method varies. The very fast \emph{get()} method
and quite fast \emph{delete()} and \emph{successor()} are contrasted
with the average \emph{insert()}. The relationships between the results
are similar to those of \cite{Kulakowski2014acve}, except \emph{insert()},
which seems to be slower than before. The need for memory allocation,
as expected, results in performance degradation. Fortunately, however,
despite the decrease in performance, the \emph{insert()} method still
has a comparable speed to the reference \emph{ConcurrentSkipListMap}.
Noteworthy is the very good result of \emph{insert()} achieved by
\emph{LockFree9ST}. Other than for the \emph{successor()} method,
the \emph{dcvEB} array turns out to be the fastest.

The \emph{dcvEB array} concurrent processing efficiency comes from
the local nature of the synchronization mechanisms (lock objects are
spread over all the AH). However, it can be assumed that this ``locality''
may not be beneficial when the requested indices are close to each
other. Hence, it is natural to test how the structure can handle situations
when the key set is small. For the test purposes it was adopted that
$K=[0,\ldots m]$, where $m$ increases starting from $10$ and ending
at $776000$. 

\begin{figure}[h]
\begin{centering}
\includegraphics[scale=0.26]{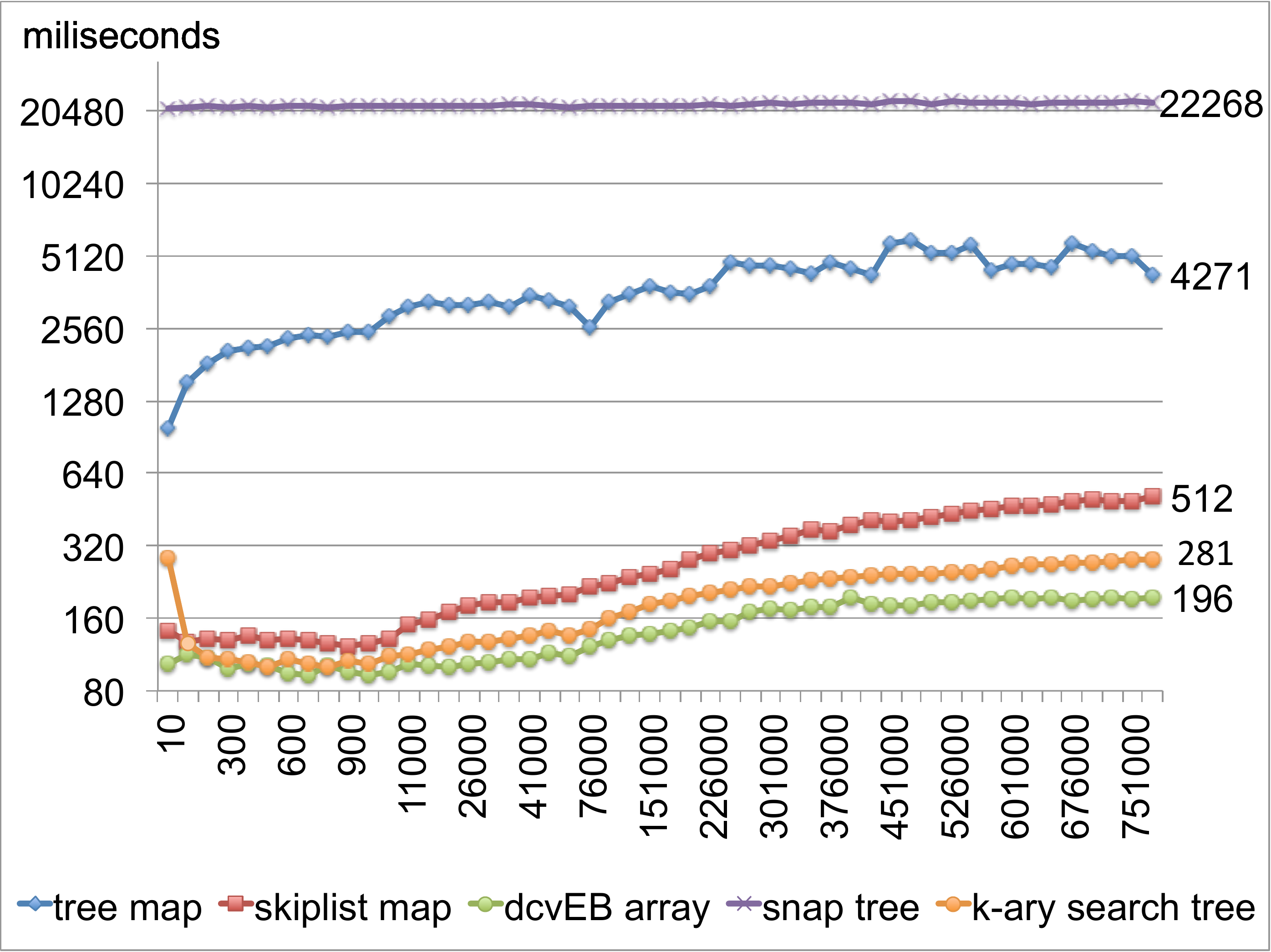}
\par\end{centering}

\caption{The \emph{dcvEB} array performance test with a variable key set ($g=i=r=s=2$,
$z=500000$ and $m=10,100,\ldots,776000$).}
\label{fig:test2}
\end{figure}

As before, for the narrow key ranges \emph{ConcurrentSkipListMap}
runs similarly fast to the \emph{dcvEB array}. The result of \emph{LockFree9ST}
does not differ significantly from those two structures. For the larger
values of $m$ the dominance of the \emph{dcvEB array} becomes apparent.
The slow increase in execution time is caused mainly by the \emph{insert()}
and \emph{successor()} methods. With the increasing key range, \emph{insert()}
needs to allocate more and more memory and the distances scanned by
\emph{successor()} get bigger. A very good result is also achieved
by \emph{LockFree9ST}. It is almost two times better than \emph{ConcurrentSkipListMap,
}although it is worse than the \emph{dcvEB array}. As in the previous
test, the results achieved by the other two structures are worse than
the results of the two ranking leaders. 

The results achieved, although encouraging, have to be interpreted
carefully. In particular, it is difficult to prejudge the performance
of the \emph{dcvEB array} relative to \emph{LockFree9ST, ConcurrentSkipListMap}
and \emph{SnapTree}. The last two of these structures have a very
powerful interface providing users with much more functionality than
the \emph{dcvEB array}. Moreover, \emph{ConcurrentSkipListMap} and
\emph{LockFree9ST} are not optimized to integer keys, hence, in return,
they can handle any object as the key. \emph{LockFree9ST }does not
support natively the \emph{successor()} and \emph{predecessor()} operations.
Hence, its actual results can be better than those observed during
the tests. Finally, \emph{SnapTree} adopts an optimistic locking strategy,
which in some cases can cause performance degradation. Despite the
better results achieved by the \emph{dcvEB array} than by the \emph{cvEB
array} \cite{Kulakowski2014acve} and the similar testing procedure,
the performance of both structures cannot be directly compared. The
presented tests were carried out on another (faster) machine than
the previous tests, the code of the testing procedure has been improved,
and another (new) version of \emph{Java} has been used. It should
also be noted that the \emph{dcvEB array} used in the experiment is
``flatter'' than the \emph{cvEB array} used in \cite{Kulakowski2014acve},
since every non-leaf node in the \emph{dcvEB array} has $64$ children,
whilst in the \emph{cvEB} array used in \cite{Kulakowski2014acve}
it was less than half of that.

\section{Comments and discussion\label{sec:Comments-and-discussion}}

The presented \emph{dcvEB array} is the result of improvements introduced
into its previous version \cite{Kulakowski2014acve}. In contrast
to the \emph{cvEB array}, the presented structure is able to grow
when more data need to be stored in it, and to shrink when the data
are removed. Hence, the amount of occupied memory (RAM) is in relation
to the amount of stored data. Of course, this relationship is not
as simple as in the case of an ordinary binary search tree or a heap
tree \cite{Cormen2009ita}. In those structures, the amount of memory
occupied depends more or less linearly on the amount of stored data.
In the case of the \emph{dcvEB} array, it also depends on the data
layout. More precisely, if the smallest key is $0$ and the largest
key is $n^{h}-1$ (where $n$ is the length of the \emph{AHs'} summary
bit vector), then the \emph{dcvEB array} can have up to $1+n+\ldots+n^{h-1}=\frac{n^{h}-1}{n-1}$
non-leaf\emph{ AHs, }and up to $n^{h}-1$ leaf \emph{AHs}. Whilst
the number of leaf \emph{AHs} depends only on the amount of data actually
stored in the \emph{dcvEB array}, the number of non-leaf AHs also
depends on data distribution and the value of the maximal key. If
the stored keys are scattered throughout the array, so that at all
$64$ positions (assuming that $n=64$) at least one element can be
found, then the memory consumption by the \emph{dcvEB array} is relatively
high. In fact, in the worst case scenario even \emph{64} times more
items could be stored in the \emph{dcvEB array} without increasing
the number of non-leaf \emph{AHs}. Reversely, if the data is indexed
by the keys arranged one after the other, memory utilization is optimal
and any increase in the amount of stored data by at least \emph{64}
keys, triggers the creation of at least one non-leaf \emph{AH}. 

Scattered keys, although undesirable due to the use of extra RAM,
may by beneficial to the methods using blocking synchronization, such
as \emph{delete()} or \emph{insert()}. The keys' dispersion increases
the chance that the threads operating on the \emph{dcvEB array} less
frequently block each other, which results in an increase in the speed
of the structure.

\section{Summary\label{sec:Summary} }

In this paper, the author proposes the \emph{dcvEB array} - the new
concurrent implementation of the dynamic set structure based on the
synchronization scheme proposed in \cite{Kulakowski2014acve}. The
preliminary tests conducted using a prototype Java-based implementation
of the structure seem to confirm that it maintains the high performance
of the concurrent operations. The introduced extensions are of great
practical importance. Thanks to them, the structure is able to dynamically
adapt its size to the amount of stored data and increase or decrease
the range of indices stored. There are also some smaller improvements,
such as a new successor search strategy. 

The variable amount of memory used, and no limit on the size of the
key in combination with the high performance of concurrent applications,
can make this structure useful for a wide range of professionals involved
in concurrent or parallel programming.

\section*{Acknowledgment}

The author would like to thank Prof. Jacek Kitowski and Piotr Matyasik,
Ph.D. (AGH) for providing the test environments, and Pawe\l{} Salata,
for helping the author in launching multiple variants of tests.

\bibliographystyle{IEEEtranS}
\bibliography{papers_biblio_reviewed}

\begin{thebibliography}{10}
\providecommand{\url}[1]{#1}
\csname url@samestyle\endcsname
\providecommand{\newblock}{\relax}
\providecommand{\bibinfo}[2]{#2}
\providecommand{\BIBentrySTDinterwordspacing}{\spaceskip=0pt\relax}
\providecommand{\BIBentryALTinterwordstretchfactor}{4}
\providecommand{\BIBentryALTinterwordspacing}{\spaceskip=\fontdimen2\font plus
\BIBentryALTinterwordstretchfactor\fontdimen3\font minus
  \fontdimen4\font\relax}
\providecommand{\BIBforeignlanguage}[2]{{%
\expandafter\ifx\csname l@#1\endcsname\relax
\typeout{** WARNING: IEEEtranS.bst: No hyphenation pattern has been}%
\typeout{** loaded for the language `#1'. Using the pattern for}%
\typeout{** the default language instead.}%
\else
\language=\csname l@#1\endcsname
\fi
#2}}
\providecommand{\BIBdecl}{\relax}
\BIBdecl

\bibitem{Bayer1972sbbtSIMPL}
R.~Bayer, ``{Symmetric binary B-Trees: Data structure and maintenance
  algorithms},'' \emph{Acta Informatica}, 1972.

\bibitem{Bayer1977coooSIMPL}
R.~Bayer and M.~Schkolnick, ``{Concurrency of Operations on B-Trees},''
  \emph{Acta Informatica}, 1977.

\bibitem{Bronson2010bsSIMPL}
N.~G. Bronson, J.~Casper, H.~Chafi, and K.~Olukotun, ``{A practical concurrent
  binary search tree},'' in \emph{Proceedings of 15th ACM SIGPLAN}, 2010.

\bibitem{Brown2011nbksSIMPL}
T.~Brown and J.~Helga, ``{Non-blocking K-ary Search Trees},'' in
  \emph{Proceedings of the 15th International Conference on Principles of
  Distributed Systems}, 2011.

\bibitem{Cormen2009ita}
T.~H. Cormen, C.~E. Leiserson, R.~L. Rivest, and C.~Stein, \emph{Introduction
  to Algorithms}, 3rd~ed.\hskip 1em plus 0.5em minus 0.4em\relax MIT Press,
  2009.

\bibitem{Ellen2010nbbsSIMPL}
F.~Ellen, P.~Fatourou, E.~Ruppert, and F.~van Breugel, ``{Non-blocking binary
  search trees},'' in \emph{Proceeding of the 29th ACM SIGACT-SIGOPS
  symposium}, 2010.

\bibitem{ELLIS1980csai}
C.~S. Ellis, ``{Concurrent Search and Insertion in 2-3 Trees},'' \emph{Acta
  Informatica}, vol.~14, no.~1, pp. 63--86, 1980.

\bibitem{Fraser2004plf}
K.~Fraser, ``{Practical lock-freedom},'' University of Cambridge, Tech. Rep.
  UCAM-CL-TR-579, 2004.

\bibitem{Hanke1997rbrb}
S.~Hanke, T.~Ottmann, and E.~Soisalon-Soininen, ``{Relaxed balanced red-black
  trees},'' \emph{Algorithms and Complexity}, 1997.

\bibitem{Herlihy2007asos}
M.~Herlihy, Y.~Lev, V.~Luchangco, and N.~Shavit, ``{A simple optimistic
  skiplist algorithm},'' in \emph{SIROCCO'07: Proceedings of the 14th
  international conference on Structural information and communication
  complexity}.\hskip 1em plus 0.5em minus 0.4em\relax Springer-Verlag, Jun.
  2007.

\bibitem{Herlihy2008taom}
M.~Herlihy and N.~Shavit, \emph{"The Art of Multiprocessor Programming"}.\hskip
  1em plus 0.5em minus 0.4em\relax Elsevier, 2008.

\bibitem{Wing1990lacc}
P.~M. Herlihy and J.~M. Wing, ``{Linearizability: a correctness condition for
  concurrent objects},'' \emph{ACM Transactions on Programming Languages and
  Systems}, vol.~12, 1990.

\bibitem{Howley2012anbiSIMPL}
S.~V. Howley and J.~Jones, ``A non-blocking internal binary search tree,'' in
  \emph{Proceedings of the 24th ACM symposium on Parallelism in algorithms and
  architectures}, 2012.

\bibitem{Knuth1973taoc}
D.~E. Knuth, \emph{The Art of Computer Programming, Volume I: Fundamental
  Algorithms, 2nd Edition}.\hskip 1em plus 0.5em minus 0.4em\relax
  Addison-Wesley, 1973.

\bibitem{Kulakowski2014acve}
\BIBentryALTinterwordspacing
K.~Ku{\l}akowski, ``{A concurrent van Emde Boas array as a fast and simple
  concurrent dynamic set alternative},'' \emph{Concurrency and Computation:
  Practice and Experience}, vol.~26, no.~2, pp. 360--379, 2014. [Online].
  Available: \url{http://dx.doi.org/10.1002/cpe.2995}
\BIBentrySTDinterwordspacing

\bibitem{Kung1980cmobSIMPL}
H.~T. Kung and P.~L. Lehman, ``{Concurrent manipulation of binary search
  trees},'' \emph{Trans. on Dat. Systems}, 1980.

\bibitem{Lanin1986ascbSIMPL}
V.~Lanin and D.~Shasha, ``A symmetric concurrent b-tree algorithm,'' in
  \emph{Proceedings of 1986 ACM Fall joint computer conference}, 1986.

\bibitem{Lea1999cpij}
D.~Lea, \emph{{Concurrent Programming in Java. Second Edition: Design
  Principles and Patterns}}, 2nd~ed.\hskip 1em plus 0.5em minus 0.4em\relax
  Boston, MA, USA: Addison-Wesley Longman Publishing Co., Inc., 1999.

\bibitem{Lehman1981hcSIMPL}
P.~L. Lehman and S.~Bing~Yao, ``{Efficient locking for concurrent operations on
  B-trees},'' \emph{Trans. on Dat. Systems}, 1981.

\bibitem{Mehlhorn1990bodi}
K.~Mehlhorn and S.~N\"{a}her, ``Bounded ordered dictionaries in o(loglogn) time
  and o(n) space,'' \emph{Inf. Process. Lett.}, vol.~35, no.~4, pp. 183--189,
  Aug. 1990.

\bibitem{Nurmi1996ratmSIMPL}
O.~Nurmi, E.~Soisalon-Soininen, and D.~Wood, ``{Relaxed AVL Trees, Main-Memory
  Databases, and Concurrency},'' HKUST, Tech. Rep., 1996.

\bibitem{Pugh1990cmos}
W.~Pugh, ``Concurrent maintenance of skip lists,'' College Park, MD, USA, Tech.
  Rep., 1990.

\bibitem{Samadi1976btia}
B.~Samadi, ``B-trees in a system with multiple users,'' \emph{Information
  Processing Letters}, pp. 107--112, 1976.

\bibitem{Shavit2011dsit}
N.~Shavit, ``{Data structures in the multicore age},'' \emph{Communication of
  the ACM}, 2011.

\bibitem{Boas1977poia}
P.~van Emde~Boas, ``Preserving order in a forest in less than logarithmic time
  and linear space,'' \emph{Inf. Process. Lett.}, vol.~6, no.~3, pp. 80--82,
  1977.

\end{thebibliography}

\end{document}